\documentclass[%
reprint,
superscriptaddress,
 amsmath,amssymb,
prb,
]{revtex4-1}

\usepackage{graphicx}
\usepackage{dcolumn}
\usepackage{bm}
\usepackage{braket}

\usepackage{todonotes}
\usepackage{xcolor}
\usepackage[normalem]{ulem}

\usepackage[pagebackref=false]{hyperref}

\begin{document}

\author{L.~Banszerus}
\thanks{These two authors contributed equally.}
\author{S.~M\"oller}
\thanks{These two authors contributed equally.}
\affiliation{JARA-FIT and 2nd Institute of Physics, RWTH Aachen University, 52074 Aachen, Germany,~EU}%
\affiliation{Peter Gr\"unberg Institute  (PGI-9), Forschungszentrum J\"ulich, 52425 J\"ulich,~Germany,~EU}
\author{K.~Hecker}
\author{E.~Icking}
\affiliation{JARA-FIT and 2nd Institute of Physics, RWTH Aachen University, 52074 Aachen, Germany,~EU}%
\affiliation{Peter Gr\"unberg Institute  (PGI-9), Forschungszentrum J\"ulich, 52425 J\"ulich,~Germany,~EU}
\author{K.~Watanabe}
\affiliation{Research Center for Functional Materials, 
National Institute for Materials Science, 1-1 Namiki, Tsukuba 305-0044, Japan}
\author{T.~Taniguchi}
\affiliation{International Center for Materials Nanoarchitectonics, 
National Institute for Materials Science,  1-1 Namiki, Tsukuba 305-0044, Japan}%
\author{F.~Hassler}
\affiliation{JARA-Institute for Quantum Information, RWTH Aachen University, 52056 Aachen, Germany, EU}
\author{C.~Volk}
\author{C.~Stampfer}
\email{stampfer@physik.rwth-aachen.de}
\affiliation{JARA-FIT and 2nd Institute of Physics, RWTH Aachen University, 52074 Aachen, Germany,~EU}%
\affiliation{Peter Gr\"unberg Institute  (PGI-9), Forschungszentrum J\"ulich, 52425 J\"ulich,~Germany,~EU}%

\title{Particle-hole symmetry protects spin-valley blockade in graphene quantum dots}

\date{\today}


\begin{abstract}
Particle-hole symmetry plays an important role for the characterization of topological phases in solid-state systems~\cite{Zirnbauer2021Feb}.
It is found, for example, in free-fermion systems at half filling, and it is closely related to the notion of antiparticles in relativistic field theories~\cite{Maurice1930Jan}.
In the low energy limit, graphene is a prime example of a gapless particle-hole symmetric system described by an effective Dirac equation~\cite{McCann2013Apr, CastroNeto2009Jan}, where topological phases can be understood by studying ways to open a gap by preserving (or breaking) symmetries~\cite{Haldane1988Oct,Qi2011Oct}. 
An important example is the intrinsic Kane-Mele spin-orbit gap of graphene, which leads to a lifting of the spin-valley degeneracy and renders graphene a topological insulator in a quantum spin Hall phase~\cite{Kane2005Nov}, while preserving particle-hole symmetry.
Here, we show that bilayer graphene allows realizing electron-hole double quantum-dots that exhibit nearly perfect particle-hole symmetry, where transport occurs via the creation and annihilation of single electron-hole pairs with opposite quantum numbers. Moreover, we show that this particle-hole symmetry results in a protected single-particle spin-valley blockade. The latter will allow robust spin-to-charge conversion and valley-to-charge conversion, which is essential for the operation of spin and valley qubits.
\end{abstract}
\maketitle

Carbon-based materials, such as monolayer and bilayer graphene, are interesting hosts for spin and spin-valley qubits, thanks to their  weak spin-orbit (SO) coupling \cite{Kane2005Nov,Konschuh2012Mar,Kurzmann2021Oct, Banszerus2021Sep} and weak hyperfine interaction~\cite{Wojtaszek2014Jan, Fischer2009Jun}. 
Bilayer graphene (BLG) is attracting in particular increasing attention, as it presents a  gate-tunable band gap, $E_\mathrm{g}$~\cite{McCann2013Apr, Icking2022Jul}, which can be used to  electrostatically confine charge carriers into quantum point contacts and quantum dots (QDs)~\cite{Eich2018Aug, Banszerus2020Oct, Banszerus2021Sep, Kurzmann2021Oct, Garreis2021Apr, Lee2020Mar}.  The small size of this gap (up to $\sim100$meV) allows forming ambipolar quantum dots, which is not possible in standard semiconductors~\cite{Banszerus2018Aug, Banszerus2020Mar, Tong2021Jan}. Another attracting feature is that, at low energy, charge carriers have an orbital magnetic moment caused by the finite Berry curvature~\cite{McCann2013Apr, Knothe2018Oct, Knothe2020Jun}. These orbital magnetic moments are aligned perpendicular to the BLG plane and allow to control the valley degree of freedom, as they have opposite signs for the two valleys ($K$ and $K'$) and for electrons and holes, as illustrated in Fig.~1a. This property is a consequence of the particle-hole symmetry that is imprinted in the low-energy Hamiltonian of bilayer graphene, 
$$ H_{\mathrm{BLG}}=-\frac1{2m}\Psi^\dag [(p_x^2 - p_y^2) \sigma_x + 2 p_x p_y \sigma_y \tau_z ] \Psi+  \frac{E_\text{g}}{2} \Psi^\dag\sigma_z \Psi \,, $$
as well as on the intrinsic Kane-Mele SO coupling term $ H_\mathrm{SO}=\frac{1}{2}\Delta_\mathrm{SO} \Psi^\dag  \sigma_z \tau_z s_z \Psi$~\cite{Kane2005Nov, Konschuh2012Mar}. 
Here, $m=0.033\; m_e$ is the effective mass of the charge carriers in BLG with the free electron mass $m_e$, $p_i$ are momentum operators and $s_i, \tau_i, \sigma_i$ are Pauli matrices ($i=x,y,z$) acting on the spin, valley and sublattice space, respectively. 
Both $H_{\mathrm{BLG}}$ and $H_{\mathrm{SO}}$  are invariant under the particle-hole transformation $\mathsf{K}$, which  effectively flips the sublattice, the valley, and the spin indices, $\mathsf{K} \Psi^\dag \mathsf{K}^{-1} = \sigma_y \tau_x s_y \Psi$. As a consequence, the hole spectrum in BLG
mirrors the electron spectrum around the $K$ and $K'$ points.

This symmetry remains true in BLG quantum dots. In this case the orbital states are quantized and form shells with four states, which are grouped by the intrinsic spin-orbit coupling into Kramers' doublets,  $\ket{K{\uparrow}}$, $\ket{K'{\downarrow}}$ and $\ket{K'{\uparrow}}$, $\ket{K{\downarrow}}$. Every electron state in the energetically lower (higher) Kramers' pair in the conduction band has a corresponding hole state in the energetically higher (lower) Kramers' pair in the valence band, as illustrated in Fig.~1b.
This is in stark contrast to the single-particle spectrum of QDs in carbon nanotubes, where the spin-orbit coupling caused by the curvature breaks the particle-hole symmetry~\cite{Pei2012Oct,Laird2015Jul}. 
Here, we show that this symmetry is almost perfectly preserved also when electron and holes are physically separated into different quantum dots and that it leads to a very strong single-particle blockade in the transport through electron-hole double quantum dots (DQDs).

\begin{figure*}[!thb]
\centering
\includegraphics[draft=false,keepaspectratio=true,clip,width=\linewidth]{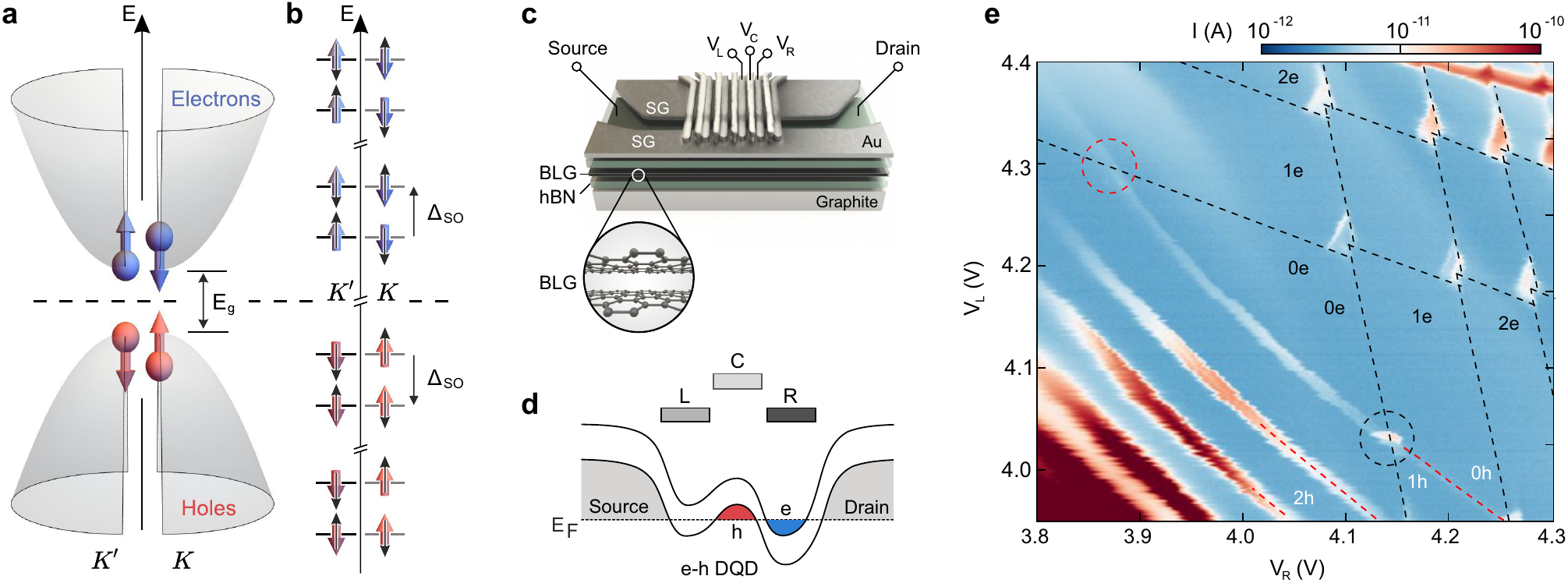}
\caption[Fig01]
{\textbf{Electron-hole symmetry in bilayer graphene and the formation of electron-hole (e-h) double QDs.}
\textbf{a} Low energy dispersion relation of gapped BLG at the $K$ and $K'$ points. The arrows indicate the orientation of the valley-dependent orbital magnetic moment of electrons (blue) and holes (red).
\textbf{b} Each orbital state in the electron and hole QDs holds four single particle states due to the spin and valley degree of freedom. The Spin-orbit gap, $\Delta_\mathrm{SO}$, splits the fourfold degeneracy of each orbital into two Kramers' pairs. Black (colored) arrows indicate the orientation of the spin (valley) magnetic moment. The first electron shell in the conduction band is separated from the first hole shell in the valence band by $E_\mathrm{g}$ and the confinement energy of the QD.
\textbf{c} Schematic cross-section of the device. The van-der-Waals heterostructure consists of a hBN/BLG/hBN/graphite stack. The gate pattern comprises a layer of split gates, forming a narrow channel, and two layers of finger gates to define and control the QDs.
\textbf{d} Schematic of the valence and conduction band edge profiles along the p-type channel. Finger gates (L, C and R) form an e-h DQD.  
\textbf{e} Charge stability diagram of the device at $V_\mathrm{SD}=1$~mV. Red dashed lines indicate charging lines of a hole QD, while black dashed lines indicate charging lines of electron QDs, with the electron (hole) occupation number labelled in black (white). The dashed circles mark the formation of single e-h DQDs. The position of the QDs along the channel is interchanged for the red and black circle.
}
\label{f1}
\end{figure*}

The DQD devices are fabricated as schematically shown in Fig.~1c. 
The devices consist of BLG encapsulated between two crystals of hexagonal boron nitride (hBN) resting on a graphite crystal that acts as a back gate.
A pair of metallic split gates (SGs) on top is used to create a narrow conducting channel (see Methods).
Two layers of interdigitated finger gates across the channel are used to modulate the band-edge profile along the channel and to confine electrons and holes in neighboring QDs, as illustrated in Fig.~\ref{f1}d~\cite{Banszerus2021Mar,Banszerus2021Sep}.
The charge stability diagram in Fig.~\ref{f1}e gives an overview of the different charge configurations of the investigated device at $V_\mathrm{SD} = 1\,$mV (see Suppl. Sec. A, for reversed bias). 
At finger gate voltages $V_\mathrm{L}$ and $V_\mathrm{R} \lesssim$ 4.1~V, a hole QD is formed underneath the central finger gate C, whereas at $V_\mathrm{L}$ and $V_\mathrm{R} \gtrsim$ 4.1~V an electron-electron DQD is formed under the left (L) and right (R) gate.
The dashed lines indicate the charge transitions of the electron (black) and hole (red) QDs.
The intersections of the electron and hole charging lines, correspond to the formation of an ambipolar electron-hole DQD.

\begin{figure*}[!thb]
\centering
\includegraphics[draft=false,keepaspectratio=true,clip,width=\linewidth]{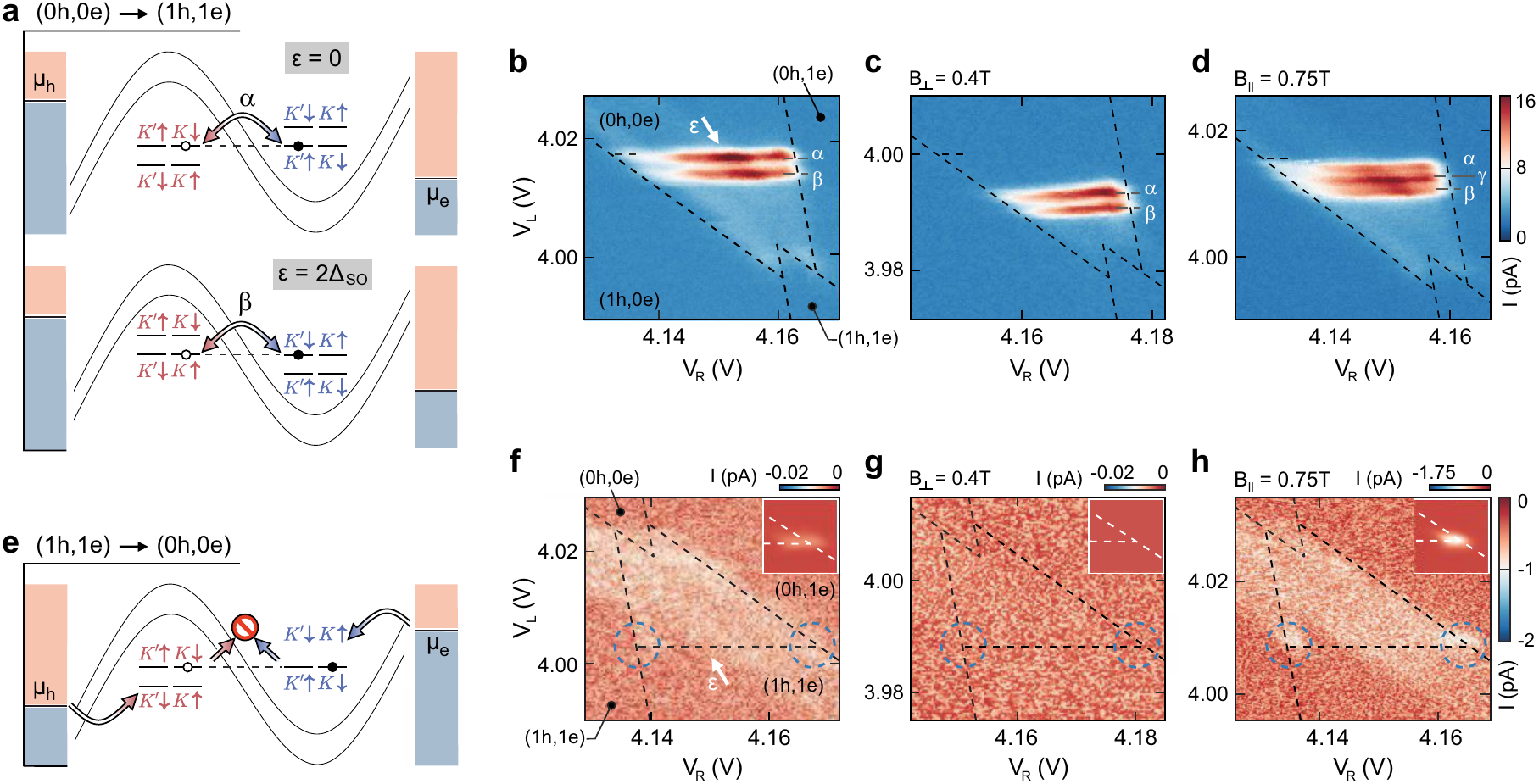}
\caption[Fig02]{\textbf{Finite bias spectroscopy and spin-valley blockade.}
\textbf{a} Schematics of the alignment of energy levels in the first shells of an electron-hole DQD for positive bias. Creation of electron-hole pairs is only possible when states of opposite spin and valley quantum numbers are aligned, hence transport only occurs via the ground state ($\alpha$) and excited state transition ($\beta$).
\textbf{b} Charge stability diagram of the $(0h,0e)\rightarrow(1h,1e)$ transition at $V_\mathrm{SD} = 1~$mV. Dashed lines show the outline of the bias triangle and co-tunneling lines, while the white arrow  indicates the direction of increasing detuning, $\varepsilon$, between the two QDs. 
\textbf{c} Charge stability diagram as in b, but for $B_\perp = 0.4$~T. The transitions $\alpha$ and $\beta$ shift with respect to the outline of the bias triangle but don't change their separation. 
\textbf{d} Charge stability diagram as in b, but for $B_\parallel = 0.75$~T. The peaks $\alpha$ and $\beta$ slightly increase their separation and an additional transition, $\gamma$, appears in between. 
\textbf{e} Schematic as in a, but for negative bias. Electrons and holes tunneling into the DQD from the leads need to recombine in order to allow for a current flow. As soon as charge carriers with incompatible spin and valley quantum number occupy the QDs, transport is blocked. 
\textbf{f-h} Charge stability diagrams as in panels b-d, but recorded at negative bias, $V_\mathrm{SD} = -1$~mV, corresponding to the $(1h,1e)\rightarrow(0h,0e)$ transition. Transport is blocked, except for faint co-tunneling effects and small currents at the corners of the bias triangles (see dashed circles). The insets show simulations of the current at the corners of the bias triangles. 
}
\label{f2}
\end{figure*}

We now focus on the triple point of the  charge transition $(0h,0e)\leftrightarrow(1h,1e)$ , highlighted by the black dashed-circle in Fig.~1e.
For positive bias voltages, a steady tunnel current through the DQD involves the transition $(0h,0e)\rightarrow(1h,1e)$, i.e. it is only possible if electron-hole pairs with opposite quantum numbers (e.g. $\ket{K{\downarrow}}_\mathrm{h}$ and $\ket{K'{\uparrow}}_\mathrm{e}$) can continuously be created. 
As the SO coupling has opposite sign for electrons and holes, this is possible only in two configurations, $\alpha$ and $\beta$, illustrated in the two panels of Fig.~2a. 
These two configurations, which are energetically offset by $2 \Delta_{\mathrm{SO}}$, result in two sharp current-peaks in the bias triangle, as shown in Fig.~\ref{f2}b. The value of $\Delta_{\mathrm{SO}}$ extracted from the separation of the peaks along the detuning axis $\Delta \varepsilon = 2\Delta_{\mathrm{SO}} = 140 \pm 10 \,\mu$eV, is in excellent agreement with the value of  $\Delta_{\mathrm{SO}}$  measured in an electron-electron DQD realized in the same device ($\Delta_{\mathrm{SO}} = 68 \pm 7 \,\mu$eV, see Suppl. Sec. B) and in other BLG QD experiments~\cite{Kurzmann2021Oct, Banszerus2020May, Banszerus2021Sep}.
The separation of the two resonances remains constant when applying a perpendicular magnetic field, only their position changes with respect to the baseline of the bias triangle, as can be seen in Fig.~2c.
In contrast, the separation between $\alpha$ and $\beta$ increases slightly when applying a parallel magnetic field, and a third resonance, $\gamma$, appears in between, as shown in Fig.~\ref{f2}d. This behavior can be well understood, as explained below. 

Before turning to this, we consider the case of negative bias voltage applied to the e-h DQD device. 
In this case, tunnel transport through the DQD requires the continuous annihilation of electron-hole pairs, i.e. $(1h,1e)\rightarrow(0h,0e)$, which is only possible if the electron and hole have opposite quantum numbers. However, since electrons and holes tunnel in from the leads with random quantum numbers, transport is blocked as soon as the QDs are occupied by charge carriers with incompatible quantum numbers, as sketched in Fig.~2e.
This blockade is robust up to very high detuning energies, because it can only be overcome by involving excited states that are separated in energy from the low energy spectrum by the band-gap, which is typically of the order of $20-60\,$meV.
The blockade is nicely visible in Fig.~\ref{f2}f, where the tunnel current inside the $(1h,1e)\rightarrow(0h,0e)$ bias triangle is entirely suppressed, except for a faint contribution that can be attributed to co-tunneling.
The current is slightly enhanced only at the corners of the bias triangle (see dashed circles), which correspond to the configurations where the electron or hole states are aligned with the Fermi level of source or drain, respectively. 
Under this condition, incompatible charge-carriers can tunnel back into the leads and new ones, possibly with matching quantum numbers, can enter the QD, lifting the blockade and allowing a small current through the DQD.
This is confirmed by measurements on another DQD (Suppl. Sec. C) and by simulations based on the single particle spectrum of BLG QDs and Pauli's master equation~\cite{Bonet2002Jan, Knothe2022Apr} (Suppl. Sec. D), which shows the lifting of the blockade only at the corners of the bias triangle, as shown in the inset of Fig.~2f-h.
Applying perpendicular magnetic fields leaves the blockade intact and reduces the co-tunneling current below the noise floor~\cite{Banszerus2021Sep,Moller2021Dec}, as shown in Fig.~2g. 
The difference between the average current inside the bias triangle and the co-tunneling current outside is below 10~fA, i.e not measurable. 
For parallel magnetic fields, the blockade also remains intact, but with a stronger enhancement of the tunnel current at the corners of the triple point, which is in agreement with the simulation (see Fig.~2h).

\begin{figure*}[!thb]
\centering
\includegraphics[draft=false,keepaspectratio=true,clip,width=\linewidth]{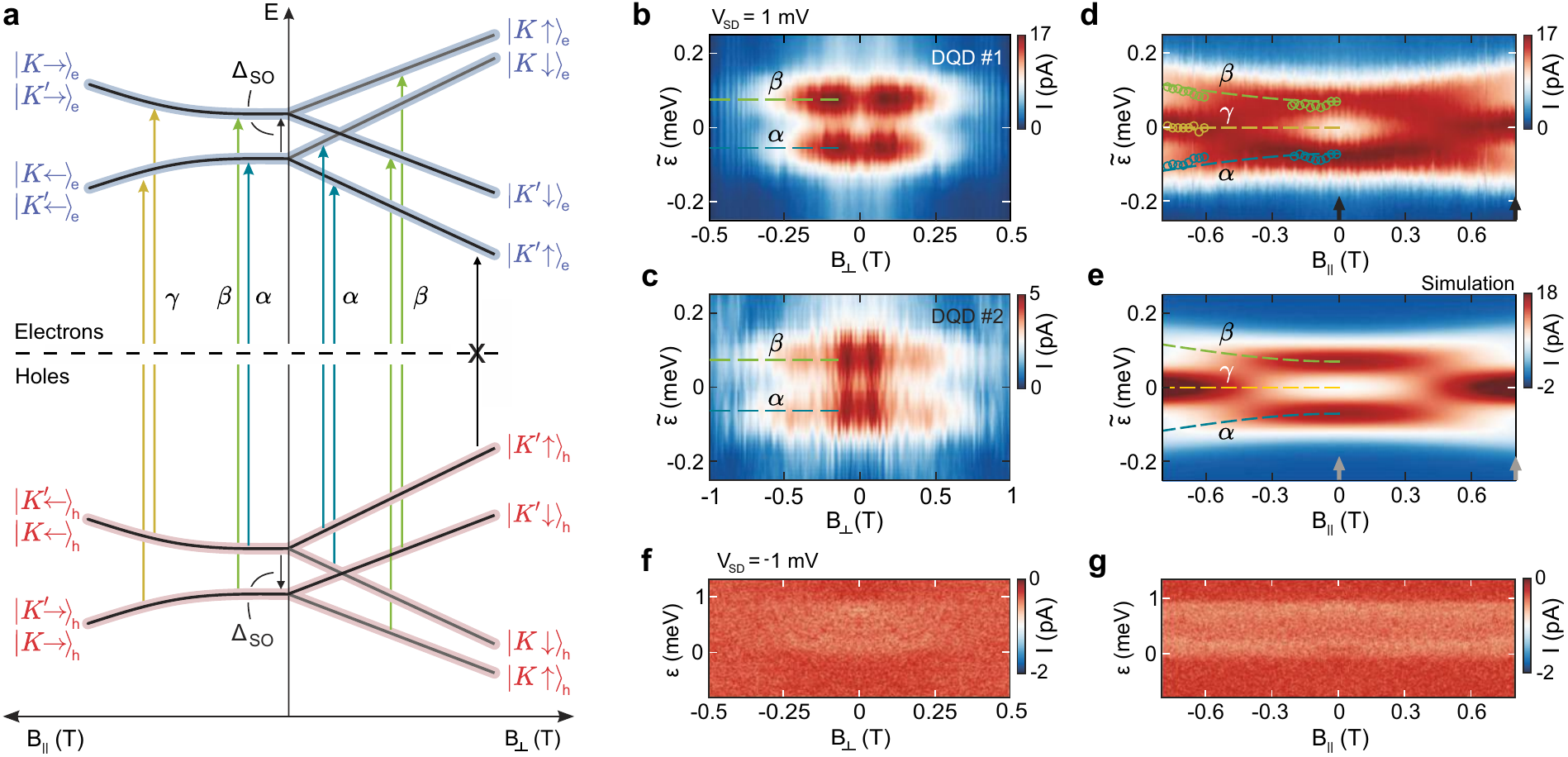}
\caption[Fig03]{\textbf{Probing the single-particle electron-hole symmetric spectrum.}
\textbf{a}~Energy dispersion of the first four single-particle electron (blue) and hole (red) states in parallel and perpendicular magnetic fields.
The Kane-Mele spin-orbit splitting, which has opposite sign for the valence and conduction band, polarizes the spins out-of-plane for zero magnetic field.
The colored lines show the required detuning of allowed transitions (creation or annihilation) between hole states in the left QD and electron states in the right QD.
\textbf{b} Current through the DQD for positive bias, $V_\mathrm{SD}=1$~mV, as function of $B_\perp$ and of detuning. 
The direction of the detuning axis is indicated by the white arrow in Fig.~\ref{f2}b. Since the position of the resonances $\alpha $ and $\beta$ shifts in gate-space as a function of $B_\perp$ (see e.g. Fig.~\ref{f2}b,c), we plot here the current as function of an effective detuning $\Tilde{\varepsilon}$, defined such that for each value of $B_\perp$ the position of the  resonances $\alpha$ and $\beta$ is symmetric with respect to $\Tilde{\varepsilon}=0$. 
\textbf{c} Same measurement as in b, but for a second DQD (see Suppl. Fig.~S3-S5).
\textbf{d} Current through the DQD as a function of $\widetilde \varepsilon$ and $B_\parallel$ for positive voltage. Colored circles indicate peak positions of the current. The colored lines indicate the expected positions of the local maxima, as given by the required detuning for each transition (c.f. length of the  arrows in a). 
\textbf{e}
Simulation of the current as a function of $\widetilde \varepsilon$ and $B_\parallel$ based on a Pauli's master equation.
\textbf{f} Current through the DQD as a function of $ \varepsilon$ and $B_\perp$ for negative bias, $V_\mathrm{SD}=-1$~mV.
\textbf{g} Same as in panel f,  but for $B_\parallel$.
}

\label{f3}
\end{figure*}

The picture of the symmetry-protected valley blockade presented above is based on the careful analysis of all possible transitions between single-hole and single-electron states in the left and right QD, and on magneto-transport measurements that support the assignment of the states involved into the various transport processes.
The magnetic-field dependent energy dispersion of the first hole and electron states is depicted in Fig.~\ref{f3}a.
When applying a perpendicular magnetic field, $B_\perp$, the degeneracy of the Kramers' pairs is lifted, and each state shifts due to the spin and valley Zeeman effect $\Delta E(B_\perp) = \frac{1}{2} (\pm g_\mathrm{s} \pm g_\mathrm{v}) \mu_\mathrm{B} B_{\perp}$. 
Here, $\mu_\mathrm{B}$ is the Bohr magneton, $g_\mathrm{s} \approx 2$ the spin g-factor  \cite{Mani2012Aug, Sichau2019Feb, Lyon2017Aug} and $g_\mathrm{v}$ the valley g-factor, which quantifies the strength of the Berry curvature induced valley-dependent orbital magnetic moment~\cite{Knothe2020Jun}. 
From Fig.~2c and Suppl.~Fig.~S5b, we extract $g_\mathrm{v} \approx 15$ for our DQD system.
Electrons and holes with opposite quantum numbers experience the same "Zeeman shift", and therefore, the  splitting between the $\alpha$ and $\beta$ transitions remains constant with $B_\perp$. 
This is clearly reflected in the magneto-transport measurements presented in Fig.~3b,c, which show the current measured along the detuning axis $\varepsilon$ of the $(0h,0e)\rightarrow(1h,1e)$ triple point (see arrow in Fig.~\ref{f2}b) as a function of $B_\perp$.

The situation is different for in-plane magnetic fields, $B_\parallel$, where the spin-Zeeman effect competes with the SO coupling, which polarizes the spins out-of-plane for zero $B$-field~\cite{Konschuh2012Mar}. With increasing $B_\parallel$, the spins are tilted into the plane of the BLG, aiming for the same spin direction within a Kramers' pair, and for opposite spin directions in different Kramers' pairs (see Fig.~\ref{f3}a, left side).
The energy difference between  Kramers' pairs increases according to $\Delta E(B_\parallel)=\pm\frac{1}{2}\sqrt{\Delta^2_\mathrm{SO}+(g_\mathrm{s} \mu_\mathrm{B} B_\parallel)^2}$.
This means that the required detuning for the $\alpha$ transition decreases, while the one for the $\beta$ transition increases, as  can be seen in Fig.~3d.
Both $\alpha$ and $\beta$ eventually vanish for $B_\parallel > 0.4\,$T, as the overlap between states in different Kramers' pairs is reduced with the increasing tilt of the spins into the plane of the BLG.
Instead, transitions involving states from the same Kramers' pair of electron and hole states become possible, i.e. pair-creation of the form $\ket{K\leftarrow}_\mathrm{e} \leftrightarrow \ket{K'\rightarrow}_\mathrm{h}$ or $\ket{K'\rightarrow}_\mathrm{e} \leftrightarrow \ket{K\leftarrow}_\mathrm{h}$. They appear as a third resonance,  $\gamma$, which is nicely visible in Fig.~3d. 
This behavior is confirmed by our simulation, assuming the energy spectrum in Fig.~\ref{f3}a with $\Delta_\mathrm{SO} = 70\,\mu$eV, $g_\mathrm{v} = 15$ and $g_\mathrm{s} = 2$. 

We confirm the robustness of the blockade for both perpendicular (Fig.~3f) and parallel (Fig.~3g) magnetic fields by repeating the measurements of Figs.~3b,d for negative bias. 
As argued above, the robustness is a direct consequence of the absence of energetically accessible excited states and of the particle-hole symmetry manifested in the spin and valley texture of our system, which forbids ground-state to ground-state transitions.
This is in stark contrast to the usual singlet-triplet Pauli-blockade in conventional semiconductors~\cite{Johnson2005Oct, Borselli2011Aug}, which can be lifted by tunneling via excited states.
The robustness of the blockade also indicates that (i) spin- or valley-flipping tunnel processes are negligible, and (ii) there is no mixing between the states within a Kramers' pair.

\begin{figure}[t]
\includegraphics[draft=false,keepaspectratio=true,clip,width=\columnwidth]{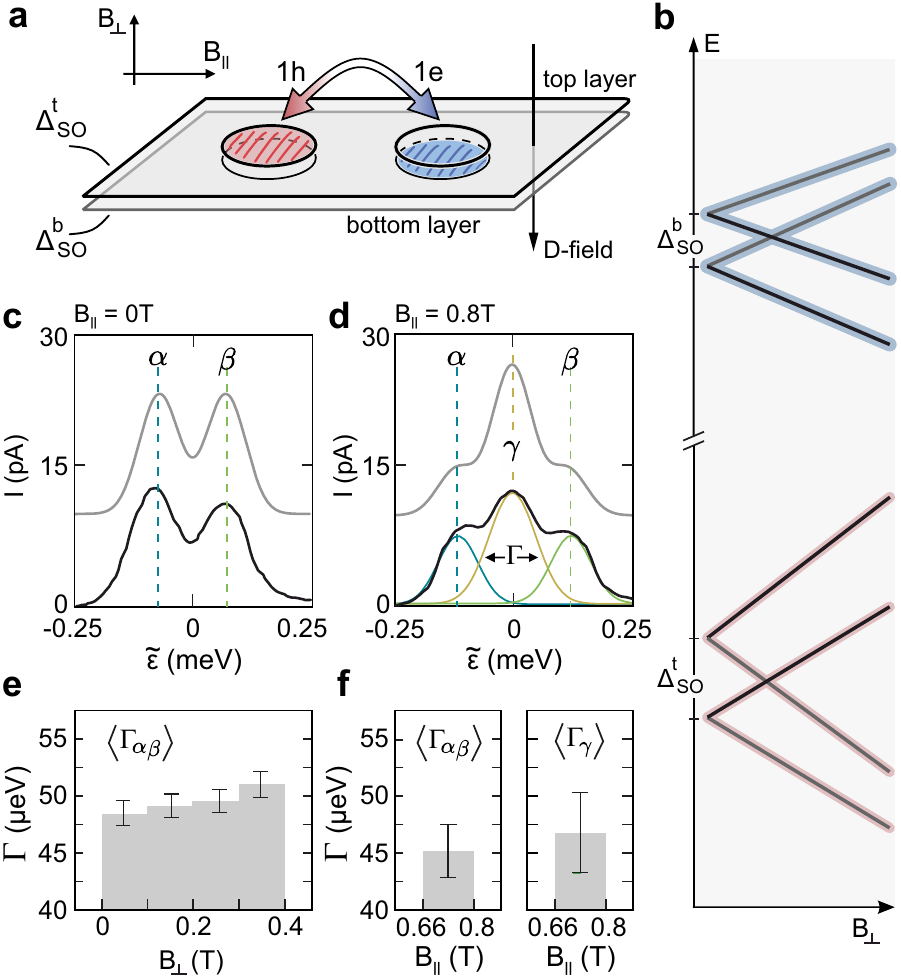}
\caption[Fig04]{ {\bf Quantitative assessment of the electron-hole symmetry}
\textbf{a} Schematic of the electron and hole QDs located on the two different layers of the BLG sheet, potentially experiencing different proximity-enhanced SO couplings on each layer. The black arrow indicates the direction of the applied displacement field.
\textbf{b} Energy dispersion of the first four single-particle electron (blue) and hole (red) states as a function of $B_\perp$ assuming $g_\mathrm{v}^\mathrm{e} \neq g_\mathrm{v}^\mathrm{h}$ and $\Delta^\text{t}_\mathrm{SO} \neq \Delta^\text{b}_\mathrm{SO}$. 
\textbf{c} Line cuts at $B_\perp=B_\parallel = 0$ along the black and gray arrows in Figs.~\ref{f3}d and~\ref{f3}e.
\textbf{d} Line cuts at $B_\parallel = 0.8$~T along the arrows in Fig.~\ref{f3}d and e. The peaks $\alpha, \beta, \gamma$ are fitted  to Gaussian line shapes. 
\textbf{e} Average width $\langle \Gamma_{\alpha\beta} \rangle$ of the resonances $\alpha$ and $\beta$ for different ranges of $B_\perp$.  Error bars indicate the 80\% percentile of $\Gamma_{\alpha\beta}$.
\textbf{f} As in e, but for $\langle \Gamma_{\alpha\beta} \rangle$ and $\langle \Gamma_{\gamma} \rangle$ at $B_\parallel = 0.66 - 0.8\,$T, where all three resonances are well distinguishable.
}
\label{f4}
\end{figure}

The experimental data are fully consistent with a description of the electron-hole DQD that is particle-hole symmetric. It should be noticed, however, that such a symmetry is not a priori granted in the system, since electrons and holes are physically separated into two different QDs on different layers of the BLG, and that the inversion symmetry is broken due to the electric displacement field, as illustrated by the schematic in Fig.~\ref{f4}a. Note that that the displacement field induced extrinsic Rashba SO coupling is not relevant for breaking the electron-hole symmetry, as discussed in Suppl. Sec. E.  There are, however, other two possible symmetry breaking mechanisms.
The first one is a difference in the valley g-factors for electrons and holes, which might be possible since the valley g-factor sensibly depends on the geometry of the QD \cite{Knothe2020Jun,Tong2021Jan,Knothe2022Apr} and electrons and holes sit in two different QDs.  
Different valley g-factors, i.e. $ g_\mathrm{v}^\mathrm{e} \neq g_\mathrm{v}^\mathrm{h} $, would break the electron-hole symmetry of the system at finite $B_\perp$ (see Fig.~\ref{f4}b), and lead to a splitting of the $\alpha$ and $\beta$ resonances with increasing $B_\perp$ (see Suppl. Sec. F). 
The second mechanism that could break the electron-hole symmetry is a difference in SO coupling for electrons and holes, $\Delta^\text{t}_\mathrm{SO} \neq \Delta^\text{b}_\mathrm{SO}$. This could originate from the fact that at low $k$-values electrons and holes are located on the different layers of BLG~\cite{McCann2013Apr,Banszerus2020May}. Thus, they can experience a different proximity-enhanced SO coupling, caused by varying inter-atomic distances or crystallographic orientations between BLG and the top and bottom hBN crystal. 
Assuming different SO coupling energies in the top~(t) and bottom~(b) layer (see Fig.~\ref{f4}a,b), the  Kane-Mele spin-orbit Hamiltonian of BLG takes then the form \cite{Banszerus2020May}
\begin{equation*}
\begin{split}
H_\mathrm{SO} =   \frac{1}{4}  \Psi^\dag\left[\left(\Delta^\text{t}_\mathrm{SO} + \Delta^\text{b}_\mathrm{SO}\right) \sigma_z - \left(\Delta^\text{t}_\mathrm{SO} -  \Delta^\text{b}_\mathrm{SO}\right) \sigma_0 \right] \tau_z s_z\Psi,  
\end{split}
\end{equation*}
which breaks the layer symmetry. Such a layer dependent SO coupling would cause a splitting of the $\gamma$ transition with a separation proportional to the asymmetry of the SO coupling between the two layers $|\Delta_\mathrm{SO}^\mathrm{t}-\Delta_\mathrm{SO}^\mathrm{b}|$.

To quantify these effects, we extract the full width at half maximum ($\Gamma$) of the resonances $\alpha$, $\beta$, and $\gamma$ by fitting Gaussian line shapes and assuming a constant background and an equal width of the $\alpha$ and $\beta$ peaks (see Fig.~\ref{f4}c,d). 
For increasing $B_\perp$, we observe a slight broadening of the $\alpha$ and $\beta$ resonance, as shown in Fig.~4e.
Attributing this broadening entirely to a difference of valley g-factors between the electron and hole QD, we get an upper limit for the valley g-factor mismatch below 1\% of $g_\text{v}$, which is consistent with the high symmetry of our gate design. 
For parallel magnetic fields, Fig.~4f shows that the average width $\langle \Gamma_{\alpha \beta} \rangle$ is comparable with $\langle \Gamma_{\gamma} \rangle$, indicating very similar SO couplings in the two layers.
From the uncertainty of the line widths, we estimate the layer asymmetry of the Kane-Mele SO coupling to be $|\Delta^\text{t}_\mathrm{SO} - \Delta^\text{b}_\mathrm{SO}| < 5\, \mu$eV, i.e below 10\% of $\Delta_\mathrm{SO}$. 

This careful analysis confirms that the system is very close to perfect particle-hole symmetric, with a rich and well-understood single-particle spectrum, in full agreement with the topologically non-trivial Kane-Mele model of a quantum spin Hall insulator.
Moreover, we demonstrated that the electron-hole symmetry leads to a strong single-particle spin and valley blockade, which is robust up to high detuning energies. 
This is in contrast to singlet-triplet Pauli spin blockade typically observed in conventional semiconductors~\cite{Johnson2005Oct, Borselli2011Aug} (including BLG~\cite{Tong2022Feb}), where the blockade is restricted to detuning energies below the singlet-triplet splitting, which can be limited by a finite valley splitting as e.g. in silicon~\cite{Borselli2011Aug}. 
Furthermore, at finite magnetic fields, the electron-hole symmetry protected blockade in BLG cannot be lifted by spin and valley relaxation, as observed in GaAs~\cite{Johnson2005Oct}.
The symmetry protected spin-valley blockade mechanism in BLG allows for spin-to-charge and valley-to-charge conversion, making it a promising read-out mechanism for spin and valley qubits.
\newline
\newline
\textbf{Methods}\\
\textbf{Sample fabrication.}
The devices are fabricated from mechnically exfoliated BLG flakes encapsulated between two hBN crystals of approximately 25~nm thickness using conventional van-der-Waals stacking techniques. A graphite flake is used as a BG. Cr/Au SGs with a lateral separation of 150~nm are deposited on top of the heterostructure. Isolated from the SGs by 15~nm thick atomic layer deposited Al$_2$O$_3$, we fabricate two layers of 70~nm wide FGs with a pitch of 150~nm. Details of the fabrication process can be found in Ref.~\cite{Banszerus2020Oct}.
\\
\textbf{Measurement technique.}
All measurements are performed in a dilution refrigerator at a base temperature of $10$~mK, using standard DC measurement techniques.
QDs are created following previous studies of gate-defined BLG QDs~\cite{Eich2018Aug,Tong2021Jan,Banszerus2021Mar}. 
Throughout the experiment, a constant BG voltage of $V_\mathrm{BG} = -1.73~$V and a SG voltage of $V_\mathrm{SG} = 1.56~$V is applied to define a p-type channel between source and drain. The estimated band gap is around 20~meV. For better comparability, the data in Figs.~3b,c,f,g is shown symmetrically around zero magnetic field.
\newline
\newline
\textbf{Acknowledgements} \\
The authors thank F. Lentz, S. Trellenkamp and D.~Neumeier for help with sample fabrication.
This project has received funding from the European Union's Horizon 2020 research and innovation programme under grant agreement No. 881603 (Graphene Flagship) and from the European Research Council (ERC) under grant agreement No. 820254, the Deutsche Forschungsgemeinschaft (DFG, German Research Foundation) under Germany's Excellence Strategy - Cluster of Excellence Matter and Light for Quantum Computing (ML4Q) EXC 2004/1 - 390534769, through DFG (STA 1146/11-1), and by the Helmholtz Nano Facility~\cite{Albrecht2017May}. K.W. and T.T. acknowledge support from JSPS KAKENHI (Grant Numbers 19H05790, 20H00354 and 21H05233). 
\\
\textbf{Data availability}\\
The data supporting the findings are available in a Zenodo repository under accession code XXX.

\textbf{Author contributions}\\
L.B. C.V. and C.S. conceived this experiment.
 L.B., S.M., K.H. and E.I. fabricated the device, L.B., S.M. and C.V. performed the measurements and analyzed the data. S.M. and F.H. performed the simulation of the current. K.W. and  T.T.  synthesized the hBN crystals. C.V. and C.S. supervised the project. L.B., S.M.,  C.V., F.H. and C.S. wrote the manuscript with contributions from all authors. L.B. and S.M. contributed equally to this work.

\textbf{Competing interests}\\
The authors declare no competing interests.


\begin{thebibliography}{36}%
\makeatletter
\providecommand \@ifxundefined [1]{%
 \@ifx{#1\undefined}
}%
\providecommand \@ifnum [1]{%
 \ifnum #1\expandafter \@firstoftwo
 \else \expandafter \@secondoftwo
 \fi
}%
\providecommand \@ifx [1]{%
 \ifx #1\expandafter \@firstoftwo
 \else \expandafter \@secondoftwo
 \fi
}%
\providecommand \natexlab [1]{#1}%
\providecommand \enquote  [1]{``#1''}%
\providecommand \bibnamefont  [1]{#1}%
\providecommand \bibfnamefont [1]{#1}%
\providecommand \citenamefont [1]{#1}%
\providecommand \href@noop [0]{\@secondoftwo}%
\providecommand \href [0]{\begingroup \@sanitize@url \@href}%
\providecommand \@href[1]{\@@startlink{#1}\@@href}%
\providecommand \@@href[1]{\endgroup#1\@@endlink}%
\providecommand \@sanitize@url [0]{\catcode `\\12\catcode `\$12\catcode
  `\&12\catcode `\#12\catcode `\^12\catcode `\_12\catcode `\%12\relax}%
\providecommand \@@startlink[1]{}%
\providecommand \@@endlink[0]{}%
\providecommand \url  [0]{\begingroup\@sanitize@url \@url }%
\providecommand \@url [1]{\endgroup\@href {#1}{\urlprefix }}%
\providecommand \urlprefix  [0]{URL }%
\providecommand \Eprint [0]{\href }%
\providecommand \doibase [0]{http://dx.doi.org/}%
\providecommand \selectlanguage [0]{\@gobble}%
\providecommand \bibinfo  [0]{\@secondoftwo}%
\providecommand \bibfield  [0]{\@secondoftwo}%
\providecommand \translation [1]{[#1]}%
\providecommand \BibitemOpen [0]{}%
\providecommand \bibitemStop [0]{}%
\providecommand \bibitemNoStop [0]{.\EOS\space}%
\providecommand \EOS [0]{\spacefactor3000\relax}%
\providecommand \BibitemShut  [1]{\csname bibitem#1\endcsname}%
\let\auto@bib@innerbib\@empty
\bibitem [{\citenamefont {Zirnbauer}(2021)}]{Zirnbauer2021Feb}%
  \BibitemOpen
  \bibfield  {author} {\bibinfo {author} {\bibfnamefont {M.~R.}\ \bibnamefont
  {Zirnbauer}},\ }\href {\doibase 10.1063/5.0035358} {\bibfield  {journal}
  {\bibinfo  {journal} {J. Math. Phys.}\ }\textbf {\bibinfo {volume} {62}},\
  \bibinfo {pages} {021101} (\bibinfo {year} {2021})}\BibitemShut {NoStop}%
\bibitem [{\citenamefont {Maurice}(1930)}]{Maurice1930Jan}%
  \BibitemOpen
  \bibfield  {author} {\bibinfo {author} {\bibfnamefont {D.~P.~A.}\
  \bibnamefont {Maurice}},\ }\href {\doibase 10.1098/rspa.1930.0013} {\bibfield
   {journal} {\bibinfo  {journal} {Proc. R. Soc. London A}\ }\textbf {\bibinfo
  {volume} {126}},\ \bibinfo {pages} {360} (\bibinfo {year}
  {1930})}\BibitemShut {NoStop}%
\bibitem [{\citenamefont {McCann}\ and\ \citenamefont
  {Koshino}(2013)}]{McCann2013Apr}%
  \BibitemOpen
  \bibfield  {author} {\bibinfo {author} {\bibfnamefont {E.}~\bibnamefont
  {McCann}}\ and\ \bibinfo {author} {\bibfnamefont {M.}~\bibnamefont
  {Koshino}},\ }\href {\doibase 10.1088/0034-4885/76/5/056503} {\bibfield
  {journal} {\bibinfo  {journal} {Rep. Prog. Phys.}\ }\textbf {\bibinfo
  {volume} {76}},\ \bibinfo {pages} {056503} (\bibinfo {year}
  {2013})}\BibitemShut {NoStop}%
\bibitem [{\citenamefont {Castro~Neto}\ \emph {et~al.}(2009)\citenamefont
  {Castro~Neto}, \citenamefont {Guinea}, \citenamefont {Peres}, \citenamefont
  {Novoselov},\ and\ \citenamefont {Geim}}]{CastroNeto2009Jan}%
  \BibitemOpen
  \bibfield  {author} {\bibinfo {author} {\bibfnamefont {A.~H.}\ \bibnamefont
  {Castro~Neto}}, \bibinfo {author} {\bibfnamefont {F.}~\bibnamefont {Guinea}},
  \bibinfo {author} {\bibfnamefont {N.~M.~R.}\ \bibnamefont {Peres}}, \bibinfo
  {author} {\bibfnamefont {K.~S.}\ \bibnamefont {Novoselov}}, \ and\ \bibinfo
  {author} {\bibfnamefont {A.~K.}\ \bibnamefont {Geim}},\ }\href {\doibase
  10.1103/RevModPhys.81.109} {\bibfield  {journal} {\bibinfo  {journal} {Rev.
  Mod. Phys.}\ }\textbf {\bibinfo {volume} {81}},\ \bibinfo {pages} {109}
  (\bibinfo {year} {2009})}\BibitemShut {NoStop}%
\bibitem [{\citenamefont {Haldane}(1988)}]{Haldane1988Oct}%
  \BibitemOpen
  \bibfield  {author} {\bibinfo {author} {\bibfnamefont {F.~D.~M.}\
  \bibnamefont {Haldane}},\ }\href {\doibase 10.1103/PhysRevLett.61.2015}
  {\bibfield  {journal} {\bibinfo  {journal} {Phys. Rev. Lett.}\ }\textbf
  {\bibinfo {volume} {61}},\ \bibinfo {pages} {2015} (\bibinfo {year}
  {1988})}\BibitemShut {NoStop}%
\bibitem [{\citenamefont {Qi}\ and\ \citenamefont {Zhang}(2011)}]{Qi2011Oct}%
  \BibitemOpen
  \bibfield  {author} {\bibinfo {author} {\bibfnamefont {X.-L.}\ \bibnamefont
  {Qi}}\ and\ \bibinfo {author} {\bibfnamefont {S.-C.}\ \bibnamefont {Zhang}},\
  }\href {\doibase 10.1103/RevModPhys.83.1057} {\bibfield  {journal} {\bibinfo
  {journal} {Rev. Mod. Phys.}\ }\textbf {\bibinfo {volume} {83}},\ \bibinfo
  {pages} {1057} (\bibinfo {year} {2011})}\BibitemShut {NoStop}%
\bibitem [{\citenamefont {Kane}\ and\ \citenamefont
  {Mele}(2005)}]{Kane2005Nov}%
  \BibitemOpen
  \bibfield  {author} {\bibinfo {author} {\bibfnamefont {C.~L.}\ \bibnamefont
  {Kane}}\ and\ \bibinfo {author} {\bibfnamefont {E.~J.}\ \bibnamefont
  {Mele}},\ }\href {\doibase 10.1103/PhysRevLett.95.226801} {\bibfield
  {journal} {\bibinfo  {journal} {Phys. Rev. Lett.}\ }\textbf {\bibinfo
  {volume} {95}},\ \bibinfo {pages} {226801} (\bibinfo {year}
  {2005})}\BibitemShut {NoStop}%
\bibitem [{\citenamefont {Konschuh}\ \emph {et~al.}(2012)\citenamefont
  {Konschuh}, \citenamefont {Gmitra}, \citenamefont {Kochan},\ and\
  \citenamefont {Fabian}}]{Konschuh2012Mar}%
  \BibitemOpen
  \bibfield  {author} {\bibinfo {author} {\bibfnamefont {S.}~\bibnamefont
  {Konschuh}}, \bibinfo {author} {\bibfnamefont {M.}~\bibnamefont {Gmitra}},
  \bibinfo {author} {\bibfnamefont {D.}~\bibnamefont {Kochan}}, \ and\ \bibinfo
  {author} {\bibfnamefont {J.}~\bibnamefont {Fabian}},\ }\href {\doibase
  10.1103/PhysRevB.85.115423} {\bibfield  {journal} {\bibinfo  {journal} {Phys.
  Rev. B}\ }\textbf {\bibinfo {volume} {85}},\ \bibinfo {pages} {115423}
  (\bibinfo {year} {2012})}\BibitemShut {NoStop}%
\bibitem [{\citenamefont {Kurzmann}\ \emph {et~al.}(2021)\citenamefont
  {Kurzmann}, \citenamefont {Kleeorin}, \citenamefont {Tong}, \citenamefont
  {Garreis}, \citenamefont {Knothe}, \citenamefont {Eich}, \citenamefont
  {Mittag}, \citenamefont {Gold}, \citenamefont {de~Vries}, \citenamefont
  {Watanabe}, \citenamefont {Taniguchi}, \citenamefont {Fal{'}ko},
  \citenamefont {Meir}, \citenamefont {Ihn},\ and\ \citenamefont
  {Ensslin}}]{Kurzmann2021Oct}%
  \BibitemOpen
  \bibfield  {author} {\bibinfo {author} {\bibfnamefont {A.}~\bibnamefont
  {Kurzmann}}, \bibinfo {author} {\bibfnamefont {Y.}~\bibnamefont {Kleeorin}},
  \bibinfo {author} {\bibfnamefont {C.}~\bibnamefont {Tong}}, \bibinfo {author}
  {\bibfnamefont {R.}~\bibnamefont {Garreis}}, \bibinfo {author} {\bibfnamefont
  {A.}~\bibnamefont {Knothe}}, \bibinfo {author} {\bibfnamefont
  {M.}~\bibnamefont {Eich}}, \bibinfo {author} {\bibfnamefont {C.}~\bibnamefont
  {Mittag}}, \bibinfo {author} {\bibfnamefont {C.}~\bibnamefont {Gold}},
  \bibinfo {author} {\bibfnamefont {F.~K.}\ \bibnamefont {de~Vries}}, \bibinfo
  {author} {\bibfnamefont {K.}~\bibnamefont {Watanabe}}, \bibinfo {author}
  {\bibfnamefont {T.}~\bibnamefont {Taniguchi}}, \bibinfo {author}
  {\bibfnamefont {V.}~\bibnamefont {Fal{'}ko}}, \bibinfo {author}
  {\bibfnamefont {Y.}~\bibnamefont {Meir}}, \bibinfo {author} {\bibfnamefont
  {T.}~\bibnamefont {Ihn}}, \ and\ \bibinfo {author} {\bibfnamefont
  {K.}~\bibnamefont {Ensslin}},\ }\href {\doibase 10.1038/s41467-021-26149-3}
  {\bibfield  {journal} {\bibinfo  {journal} {Nat. Commun.}\ }\textbf {\bibinfo
  {volume} {12}},\ \bibinfo {pages} {6004} (\bibinfo {year}
  {2021})}\BibitemShut {NoStop}%
\bibitem [{\citenamefont {Banszerus}\ \emph
  {et~al.}(2021{\natexlab{a}})\citenamefont {Banszerus}, \citenamefont
  {M{\ifmmode\ddot{o}\else\"{o}\fi}ller}, \citenamefont {Steiner},
  \citenamefont {Icking}, \citenamefont {Trellenkamp}, \citenamefont {Lentz},
  \citenamefont {Watanabe}, \citenamefont {Taniguchi}, \citenamefont {Volk},\
  and\ \citenamefont {Stampfer}}]{Banszerus2021Sep}%
  \BibitemOpen
  \bibfield  {author} {\bibinfo {author} {\bibfnamefont {L.}~\bibnamefont
  {Banszerus}}, \bibinfo {author} {\bibfnamefont {S.}~\bibnamefont
  {M{\ifmmode\ddot{o}\else\"{o}\fi}ller}}, \bibinfo {author} {\bibfnamefont
  {C.}~\bibnamefont {Steiner}}, \bibinfo {author} {\bibfnamefont
  {E.}~\bibnamefont {Icking}}, \bibinfo {author} {\bibfnamefont
  {S.}~\bibnamefont {Trellenkamp}}, \bibinfo {author} {\bibfnamefont
  {F.}~\bibnamefont {Lentz}}, \bibinfo {author} {\bibfnamefont
  {K.}~\bibnamefont {Watanabe}}, \bibinfo {author} {\bibfnamefont
  {T.}~\bibnamefont {Taniguchi}}, \bibinfo {author} {\bibfnamefont
  {C.}~\bibnamefont {Volk}}, \ and\ \bibinfo {author} {\bibfnamefont
  {C.}~\bibnamefont {Stampfer}},\ }\href {\doibase 10.1038/s41467-021-25498-3}
  {\bibfield  {journal} {\bibinfo  {journal} {Nat. Commun.}\ }\textbf {\bibinfo
  {volume} {12}},\ \bibinfo {pages} {5250} (\bibinfo {year}
  {2021}{\natexlab{a}})}\BibitemShut {NoStop}%
\bibitem [{\citenamefont {Wojtaszek}\ \emph {et~al.}(2014)\citenamefont
  {Wojtaszek}, \citenamefont {Vera-Marun}, \citenamefont {Whiteway},
  \citenamefont {Hilke},\ and\ \citenamefont {van Wees}}]{Wojtaszek2014Jan}%
  \BibitemOpen
  \bibfield  {author} {\bibinfo {author} {\bibfnamefont {M.}~\bibnamefont
  {Wojtaszek}}, \bibinfo {author} {\bibfnamefont {I.~J.}\ \bibnamefont
  {Vera-Marun}}, \bibinfo {author} {\bibfnamefont {E.}~\bibnamefont
  {Whiteway}}, \bibinfo {author} {\bibfnamefont {M.}~\bibnamefont {Hilke}}, \
  and\ \bibinfo {author} {\bibfnamefont {B.~J.}\ \bibnamefont {van Wees}},\
  }\href {\doibase 10.1103/PhysRevB.89.035417} {\bibfield  {journal} {\bibinfo
  {journal} {Phys. Rev. B}\ }\textbf {\bibinfo {volume} {89}},\ \bibinfo
  {pages} {035417} (\bibinfo {year} {2014})}\BibitemShut {NoStop}%
\bibitem [{\citenamefont {Fischer}\ and\ \citenamefont
  {Loss}(2009)}]{Fischer2009Jun}%
  \BibitemOpen
  \bibfield  {author} {\bibinfo {author} {\bibfnamefont {J.}~\bibnamefont
  {Fischer}}\ and\ \bibinfo {author} {\bibfnamefont {D.}~\bibnamefont {Loss}},\
  }\href {\doibase 10.1126/science.1169554} {\bibfield  {journal} {\bibinfo
  {journal} {Science}\ }\textbf {\bibinfo {volume} {324}},\ \bibinfo {pages}
  {1277} (\bibinfo {year} {2009})}\BibitemShut {NoStop}%
\bibitem [{\citenamefont {Icking}\ \emph {et~al.}(2022)\citenamefont {Icking},
  \citenamefont {Banszerus}, \citenamefont
  {W{\ifmmode\ddot{o}\else\"{o}\fi}rtche}, \citenamefont {Volmer},
  \citenamefont {Schmidt}, \citenamefont {Steiner}, \citenamefont {Engels},
  \citenamefont {Hesselmann}, \citenamefont {Goldsche}, \citenamefont
  {Watanabe}, \citenamefont {Taniguchi}, \citenamefont {Volk}, \citenamefont
  {Beschoten},\ and\ \citenamefont {Stampfer}}]{Icking2022Jul}%
  \BibitemOpen
  \bibfield  {author} {\bibinfo {author} {\bibfnamefont {E.}~\bibnamefont
  {Icking}}, \bibinfo {author} {\bibfnamefont {L.}~\bibnamefont {Banszerus}},
  \bibinfo {author} {\bibfnamefont {F.}~\bibnamefont
  {W{\ifmmode\ddot{o}\else\"{o}\fi}rtche}}, \bibinfo {author} {\bibfnamefont
  {F.}~\bibnamefont {Volmer}}, \bibinfo {author} {\bibfnamefont
  {P.}~\bibnamefont {Schmidt}}, \bibinfo {author} {\bibfnamefont
  {C.}~\bibnamefont {Steiner}}, \bibinfo {author} {\bibfnamefont
  {S.}~\bibnamefont {Engels}}, \bibinfo {author} {\bibfnamefont
  {J.}~\bibnamefont {Hesselmann}}, \bibinfo {author} {\bibfnamefont
  {M.}~\bibnamefont {Goldsche}}, \bibinfo {author} {\bibfnamefont
  {K.}~\bibnamefont {Watanabe}}, \bibinfo {author} {\bibfnamefont
  {T.}~\bibnamefont {Taniguchi}}, \bibinfo {author} {\bibfnamefont
  {C.}~\bibnamefont {Volk}}, \bibinfo {author} {\bibfnamefont {B.}~\bibnamefont
  {Beschoten}}, \ and\ \bibinfo {author} {\bibfnamefont {C.}~\bibnamefont
  {Stampfer}},\ }\href {\doibase 10.1002/aelm.202200510} {\bibfield  {journal}
  {\bibinfo  {journal} {Adv. Electron. Mater.}\ }\textbf {\bibinfo {volume}
  {8}},\ \bibinfo {pages} {2200510} (\bibinfo {year} {2022})}\BibitemShut
  {NoStop}%
\bibitem [{\citenamefont {Eich}\ \emph {et~al.}(2018)\citenamefont {Eich},
  \citenamefont {Pisoni}, \citenamefont {Pally}, \citenamefont {Overweg},
  \citenamefont {Kurzmann}, \citenamefont {Lee}, \citenamefont {Rickhaus},
  \citenamefont {Watanabe}, \citenamefont {Taniguchi}, \citenamefont
  {Ensslin},\ and\ \citenamefont {Ihn}}]{Eich2018Aug}%
  \BibitemOpen
  \bibfield  {author} {\bibinfo {author} {\bibfnamefont {M.}~\bibnamefont
  {Eich}}, \bibinfo {author} {\bibfnamefont {R.}~\bibnamefont {Pisoni}},
  \bibinfo {author} {\bibfnamefont {A.}~\bibnamefont {Pally}}, \bibinfo
  {author} {\bibfnamefont {H.}~\bibnamefont {Overweg}}, \bibinfo {author}
  {\bibfnamefont {A.}~\bibnamefont {Kurzmann}}, \bibinfo {author}
  {\bibfnamefont {Y.}~\bibnamefont {Lee}}, \bibinfo {author} {\bibfnamefont
  {P.}~\bibnamefont {Rickhaus}}, \bibinfo {author} {\bibfnamefont
  {K.}~\bibnamefont {Watanabe}}, \bibinfo {author} {\bibfnamefont
  {T.}~\bibnamefont {Taniguchi}}, \bibinfo {author} {\bibfnamefont
  {K.}~\bibnamefont {Ensslin}}, \ and\ \bibinfo {author} {\bibfnamefont
  {T.}~\bibnamefont {Ihn}},\ }\href {\doibase 10.1021/acs.nanolett.8b01859}
  {\bibfield  {journal} {\bibinfo  {journal} {Nano Lett.}\ }\textbf {\bibinfo
  {volume} {18}},\ \bibinfo {pages} {5042} (\bibinfo {year}
  {2018})}\BibitemShut {NoStop}%
\bibitem [{\citenamefont {Banszerus}\ \emph
  {et~al.}(2020{\natexlab{a}})\citenamefont {Banszerus}, \citenamefont
  {Rothstein}, \citenamefont {Fabian}, \citenamefont
  {M{\ifmmode\ddot{o}\else\"{o}\fi}ller}, \citenamefont {Icking}, \citenamefont
  {Trellenkamp}, \citenamefont {Lentz}, \citenamefont {Neumaier}, \citenamefont
  {Watanabe}, \citenamefont {Taniguchi}, \citenamefont {Libisch}, \citenamefont
  {Volk},\ and\ \citenamefont {Stampfer}}]{Banszerus2020Oct}%
  \BibitemOpen
  \bibfield  {author} {\bibinfo {author} {\bibfnamefont {L.}~\bibnamefont
  {Banszerus}}, \bibinfo {author} {\bibfnamefont {A.}~\bibnamefont
  {Rothstein}}, \bibinfo {author} {\bibfnamefont {T.}~\bibnamefont {Fabian}},
  \bibinfo {author} {\bibfnamefont {S.}~\bibnamefont
  {M{\ifmmode\ddot{o}\else\"{o}\fi}ller}}, \bibinfo {author} {\bibfnamefont
  {E.}~\bibnamefont {Icking}}, \bibinfo {author} {\bibfnamefont
  {S.}~\bibnamefont {Trellenkamp}}, \bibinfo {author} {\bibfnamefont
  {F.}~\bibnamefont {Lentz}}, \bibinfo {author} {\bibfnamefont
  {D.}~\bibnamefont {Neumaier}}, \bibinfo {author} {\bibfnamefont
  {K.}~\bibnamefont {Watanabe}}, \bibinfo {author} {\bibfnamefont
  {T.}~\bibnamefont {Taniguchi}}, \bibinfo {author} {\bibfnamefont
  {F.}~\bibnamefont {Libisch}}, \bibinfo {author} {\bibfnamefont
  {C.}~\bibnamefont {Volk}}, \ and\ \bibinfo {author} {\bibfnamefont
  {C.}~\bibnamefont {Stampfer}},\ }\href {\doibase
  10.1021/acs.nanolett.0c03227} {\bibfield  {journal} {\bibinfo  {journal}
  {Nano Lett.}\ }\textbf {\bibinfo {volume} {20}},\ \bibinfo {pages} {7709}
  (\bibinfo {year} {2020}{\natexlab{a}})}\BibitemShut {NoStop}%
\bibitem [{\citenamefont {Garreis}\ \emph {et~al.}(2021)\citenamefont
  {Garreis}, \citenamefont {Knothe}, \citenamefont {Tong}, \citenamefont
  {Eich}, \citenamefont {Gold}, \citenamefont {Watanabe}, \citenamefont
  {Taniguchi}, \citenamefont {Fal{'}ko}, \citenamefont {Ihn}, \citenamefont
  {Ensslin},\ and\ \citenamefont {Kurzmann}}]{Garreis2021Apr}%
  \BibitemOpen
  \bibfield  {author} {\bibinfo {author} {\bibfnamefont {R.}~\bibnamefont
  {Garreis}}, \bibinfo {author} {\bibfnamefont {A.}~\bibnamefont {Knothe}},
  \bibinfo {author} {\bibfnamefont {C.}~\bibnamefont {Tong}}, \bibinfo {author}
  {\bibfnamefont {M.}~\bibnamefont {Eich}}, \bibinfo {author} {\bibfnamefont
  {C.}~\bibnamefont {Gold}}, \bibinfo {author} {\bibfnamefont {K.}~\bibnamefont
  {Watanabe}}, \bibinfo {author} {\bibfnamefont {T.}~\bibnamefont {Taniguchi}},
  \bibinfo {author} {\bibfnamefont {V.}~\bibnamefont {Fal{'}ko}}, \bibinfo
  {author} {\bibfnamefont {T.}~\bibnamefont {Ihn}}, \bibinfo {author}
  {\bibfnamefont {K.}~\bibnamefont {Ensslin}}, \ and\ \bibinfo {author}
  {\bibfnamefont {A.}~\bibnamefont {Kurzmann}},\ }\href {\doibase
  10.1103/PhysRevLett.126.147703} {\bibfield  {journal} {\bibinfo  {journal}
  {Phys. Rev. Lett.}\ }\textbf {\bibinfo {volume} {126}},\ \bibinfo {pages}
  {147703} (\bibinfo {year} {2021})}\BibitemShut {NoStop}%
\bibitem [{\citenamefont {Lee}\ \emph {et~al.}(2020)\citenamefont {Lee},
  \citenamefont {Knothe}, \citenamefont {Overweg}, \citenamefont {Eich},
  \citenamefont {Gold}, \citenamefont {Kurzmann}, \citenamefont {Klasovika},
  \citenamefont {Taniguchi}, \citenamefont {Wantanabe}, \citenamefont
  {Fal{'}ko}, \citenamefont {Ihn}, \citenamefont {Ensslin},\ and\ \citenamefont
  {Rickhaus}}]{Lee2020Mar}%
  \BibitemOpen
  \bibfield  {author} {\bibinfo {author} {\bibfnamefont {Y.}~\bibnamefont
  {Lee}}, \bibinfo {author} {\bibfnamefont {A.}~\bibnamefont {Knothe}},
  \bibinfo {author} {\bibfnamefont {H.}~\bibnamefont {Overweg}}, \bibinfo
  {author} {\bibfnamefont {M.}~\bibnamefont {Eich}}, \bibinfo {author}
  {\bibfnamefont {C.}~\bibnamefont {Gold}}, \bibinfo {author} {\bibfnamefont
  {A.}~\bibnamefont {Kurzmann}}, \bibinfo {author} {\bibfnamefont
  {V.}~\bibnamefont {Klasovika}}, \bibinfo {author} {\bibfnamefont
  {T.}~\bibnamefont {Taniguchi}}, \bibinfo {author} {\bibfnamefont
  {K.}~\bibnamefont {Wantanabe}}, \bibinfo {author} {\bibfnamefont
  {V.}~\bibnamefont {Fal{'}ko}}, \bibinfo {author} {\bibfnamefont
  {T.}~\bibnamefont {Ihn}}, \bibinfo {author} {\bibfnamefont {K.}~\bibnamefont
  {Ensslin}}, \ and\ \bibinfo {author} {\bibfnamefont {P.}~\bibnamefont
  {Rickhaus}},\ }\href {\doibase 10.1103/PhysRevLett.124.126802} {\bibfield
  {journal} {\bibinfo  {journal} {Phys. Rev. Lett.}\ }\textbf {\bibinfo
  {volume} {124}},\ \bibinfo {pages} {126802} (\bibinfo {year}
  {2020})}\BibitemShut {NoStop}%
\bibitem [{\citenamefont {Banszerus}\ \emph {et~al.}(2018)\citenamefont
  {Banszerus}, \citenamefont {Frohn}, \citenamefont {Epping}, \citenamefont
  {Neumaier}, \citenamefont {Watanabe}, \citenamefont {Taniguchi},\ and\
  \citenamefont {Stampfer}}]{Banszerus2018Aug}%
  \BibitemOpen
  \bibfield  {author} {\bibinfo {author} {\bibfnamefont {L.}~\bibnamefont
  {Banszerus}}, \bibinfo {author} {\bibfnamefont {B.}~\bibnamefont {Frohn}},
  \bibinfo {author} {\bibfnamefont {A.}~\bibnamefont {Epping}}, \bibinfo
  {author} {\bibfnamefont {D.}~\bibnamefont {Neumaier}}, \bibinfo {author}
  {\bibfnamefont {K.}~\bibnamefont {Watanabe}}, \bibinfo {author}
  {\bibfnamefont {T.}~\bibnamefont {Taniguchi}}, \ and\ \bibinfo {author}
  {\bibfnamefont {C.}~\bibnamefont {Stampfer}},\ }\href {\doibase
  10.1021/acs.nanolett.8b01303} {\bibfield  {journal} {\bibinfo  {journal}
  {Nano Lett.}\ }\textbf {\bibinfo {volume} {18}},\ \bibinfo {pages} {4785}
  (\bibinfo {year} {2018})}\BibitemShut {NoStop}%
\bibitem [{\citenamefont {Banszerus}\ \emph
  {et~al.}(2020{\natexlab{b}})\citenamefont {Banszerus}, \citenamefont
  {M{\ifmmode\ddot{o}\else\"{o}\fi}ller}, \citenamefont {Icking}, \citenamefont
  {Watanabe}, \citenamefont {Taniguchi}, \citenamefont {Volk},\ and\
  \citenamefont {Stampfer}}]{Banszerus2020Mar}%
  \BibitemOpen
  \bibfield  {author} {\bibinfo {author} {\bibfnamefont {L.}~\bibnamefont
  {Banszerus}}, \bibinfo {author} {\bibfnamefont {S.}~\bibnamefont
  {M{\ifmmode\ddot{o}\else\"{o}\fi}ller}}, \bibinfo {author} {\bibfnamefont
  {E.}~\bibnamefont {Icking}}, \bibinfo {author} {\bibfnamefont
  {K.}~\bibnamefont {Watanabe}}, \bibinfo {author} {\bibfnamefont
  {T.}~\bibnamefont {Taniguchi}}, \bibinfo {author} {\bibfnamefont
  {C.}~\bibnamefont {Volk}}, \ and\ \bibinfo {author} {\bibfnamefont
  {C.}~\bibnamefont {Stampfer}},\ }\href {\doibase
  10.1021/acs.nanolett.9b05295} {\bibfield  {journal} {\bibinfo  {journal}
  {Nano Lett.}\ }\textbf {\bibinfo {volume} {20}},\ \bibinfo {pages} {2005}
  (\bibinfo {year} {2020}{\natexlab{b}})}\BibitemShut {NoStop}%
\bibitem [{\citenamefont {Tong}\ \emph {et~al.}(2021)\citenamefont {Tong},
  \citenamefont {Garreis}, \citenamefont {Knothe}, \citenamefont {Eich},
  \citenamefont {Sacchi}, \citenamefont {Watanabe}, \citenamefont {Taniguchi},
  \citenamefont {Fal{'}ko}, \citenamefont {Ihn}, \citenamefont {Ensslin},\ and\
  \citenamefont {Kurzmann}}]{Tong2021Jan}%
  \BibitemOpen
  \bibfield  {author} {\bibinfo {author} {\bibfnamefont {C.}~\bibnamefont
  {Tong}}, \bibinfo {author} {\bibfnamefont {R.}~\bibnamefont {Garreis}},
  \bibinfo {author} {\bibfnamefont {A.}~\bibnamefont {Knothe}}, \bibinfo
  {author} {\bibfnamefont {M.}~\bibnamefont {Eich}}, \bibinfo {author}
  {\bibfnamefont {A.}~\bibnamefont {Sacchi}}, \bibinfo {author} {\bibfnamefont
  {K.}~\bibnamefont {Watanabe}}, \bibinfo {author} {\bibfnamefont
  {T.}~\bibnamefont {Taniguchi}}, \bibinfo {author} {\bibfnamefont
  {V.}~\bibnamefont {Fal{'}ko}}, \bibinfo {author} {\bibfnamefont
  {T.}~\bibnamefont {Ihn}}, \bibinfo {author} {\bibfnamefont {K.}~\bibnamefont
  {Ensslin}}, \ and\ \bibinfo {author} {\bibfnamefont {A.}~\bibnamefont
  {Kurzmann}},\ }\href {\doibase 10.1021/acs.nanolett.0c04343} {\bibfield
  {journal} {\bibinfo  {journal} {Nano Lett.}\ }\textbf {\bibinfo {volume}
  {21}},\ \bibinfo {pages} {1068} (\bibinfo {year} {2021})}\BibitemShut
  {NoStop}%
\bibitem [{\citenamefont {Knothe}\ and\ \citenamefont
  {Fal'ko}(2018)}]{Knothe2018Oct}%
  \BibitemOpen
  \bibfield  {author} {\bibinfo {author} {\bibfnamefont {A.}~\bibnamefont
  {Knothe}}\ and\ \bibinfo {author} {\bibfnamefont {V.}~\bibnamefont
  {Fal'ko}},\ }\href {\doibase 10.1103/PhysRevB.98.155435} {\bibfield
  {journal} {\bibinfo  {journal} {Phys. Rev. B}\ }\textbf {\bibinfo {volume}
  {98}},\ \bibinfo {pages} {155435} (\bibinfo {year} {2018})}\BibitemShut
  {NoStop}%
\bibitem [{\citenamefont {Knothe}\ and\ \citenamefont
  {Fal'ko}(2020)}]{Knothe2020Jun}%
  \BibitemOpen
  \bibfield  {author} {\bibinfo {author} {\bibfnamefont {A.}~\bibnamefont
  {Knothe}}\ and\ \bibinfo {author} {\bibfnamefont {V.}~\bibnamefont
  {Fal'ko}},\ }\href {\doibase 10.1103/PhysRevB.101.235423} {\bibfield
  {journal} {\bibinfo  {journal} {Phys. Rev. B}\ }\textbf {\bibinfo {volume}
  {101}},\ \bibinfo {pages} {235423} (\bibinfo {year} {2020})}\BibitemShut
  {NoStop}%
\bibitem [{\citenamefont {Pei}\ \emph {et~al.}(2012)\citenamefont {Pei},
  \citenamefont {Laird}, \citenamefont {Steele},\ and\ \citenamefont
  {Kouwenhoven}}]{Pei2012Oct}%
  \BibitemOpen
  \bibfield  {author} {\bibinfo {author} {\bibfnamefont {F.}~\bibnamefont
  {Pei}}, \bibinfo {author} {\bibfnamefont {E.~A.}\ \bibnamefont {Laird}},
  \bibinfo {author} {\bibfnamefont {G.~A.}\ \bibnamefont {Steele}}, \ and\
  \bibinfo {author} {\bibfnamefont {L.~P.}\ \bibnamefont {Kouwenhoven}},\
  }\href {\doibase 10.1038/nnano.2012.160} {\bibfield  {journal} {\bibinfo
  {journal} {Nat. Nanotechnol.}\ }\textbf {\bibinfo {volume} {7}},\ \bibinfo
  {pages} {630} (\bibinfo {year} {2012})}\BibitemShut {NoStop}%
\bibitem [{\citenamefont {Laird}\ \emph {et~al.}(2015)\citenamefont {Laird},
  \citenamefont {Kuemmeth}, \citenamefont {Steele}, \citenamefont
  {Grove-Rasmussen}, \citenamefont {Nyg{\aa}rd}, \citenamefont {Flensberg},\
  and\ \citenamefont {Kouwenhoven}}]{Laird2015Jul}%
  \BibitemOpen
  \bibfield  {author} {\bibinfo {author} {\bibfnamefont {E.~A.}\ \bibnamefont
  {Laird}}, \bibinfo {author} {\bibfnamefont {F.}~\bibnamefont {Kuemmeth}},
  \bibinfo {author} {\bibfnamefont {G.~A.}\ \bibnamefont {Steele}}, \bibinfo
  {author} {\bibfnamefont {K.}~\bibnamefont {Grove-Rasmussen}}, \bibinfo
  {author} {\bibfnamefont {J.}~\bibnamefont {Nyg{\aa}rd}}, \bibinfo {author}
  {\bibfnamefont {K.}~\bibnamefont {Flensberg}}, \ and\ \bibinfo {author}
  {\bibfnamefont {L.~P.}\ \bibnamefont {Kouwenhoven}},\ }\href {\doibase
  10.1103/RevModPhys.87.703} {\bibfield  {journal} {\bibinfo  {journal} {Rev.
  Mod. Phys.}\ }\textbf {\bibinfo {volume} {87}},\ \bibinfo {pages} {703}
  (\bibinfo {year} {2015})}\BibitemShut {NoStop}%
\bibitem [{\citenamefont {Banszerus}\ \emph
  {et~al.}(2021{\natexlab{b}})\citenamefont {Banszerus}, \citenamefont
  {Rothstein}, \citenamefont {Icking}, \citenamefont
  {M{\ifmmode\ddot{o}\else\"{o}\fi}ller}, \citenamefont {Watanabe},
  \citenamefont {Taniguchi}, \citenamefont {Stampfer},\ and\ \citenamefont
  {Volk}}]{Banszerus2021Mar}%
  \BibitemOpen
  \bibfield  {author} {\bibinfo {author} {\bibfnamefont {L.}~\bibnamefont
  {Banszerus}}, \bibinfo {author} {\bibfnamefont {A.}~\bibnamefont
  {Rothstein}}, \bibinfo {author} {\bibfnamefont {E.}~\bibnamefont {Icking}},
  \bibinfo {author} {\bibfnamefont {S.}~\bibnamefont
  {M{\ifmmode\ddot{o}\else\"{o}\fi}ller}}, \bibinfo {author} {\bibfnamefont
  {K.}~\bibnamefont {Watanabe}}, \bibinfo {author} {\bibfnamefont
  {T.}~\bibnamefont {Taniguchi}}, \bibinfo {author} {\bibfnamefont
  {C.}~\bibnamefont {Stampfer}}, \ and\ \bibinfo {author} {\bibfnamefont
  {C.}~\bibnamefont {Volk}},\ }\href {\doibase 10.1063/5.0035300} {\bibfield
  {journal} {\bibinfo  {journal} {Appl. Phys. Lett.}\ }\textbf {\bibinfo
  {volume} {118}},\ \bibinfo {pages} {103101} (\bibinfo {year}
  {2021}{\natexlab{b}})}\BibitemShut {NoStop}%
\bibitem [{\citenamefont {Banszerus}\ \emph
  {et~al.}(2020{\natexlab{c}})\citenamefont {Banszerus}, \citenamefont {Frohn},
  \citenamefont {Fabian}, \citenamefont {Somanchi}, \citenamefont {Epping},
  \citenamefont {M{\ifmmode\ddot{u}\else\"{u}\fi}ller}, \citenamefont
  {Neumaier}, \citenamefont {Watanabe}, \citenamefont {Taniguchi},
  \citenamefont {Libisch}, \citenamefont {Beschoten}, \citenamefont {Hassler},\
  and\ \citenamefont {Stampfer}}]{Banszerus2020May}%
  \BibitemOpen
  \bibfield  {author} {\bibinfo {author} {\bibfnamefont {L.}~\bibnamefont
  {Banszerus}}, \bibinfo {author} {\bibfnamefont {B.}~\bibnamefont {Frohn}},
  \bibinfo {author} {\bibfnamefont {T.}~\bibnamefont {Fabian}}, \bibinfo
  {author} {\bibfnamefont {S.}~\bibnamefont {Somanchi}}, \bibinfo {author}
  {\bibfnamefont {A.}~\bibnamefont {Epping}}, \bibinfo {author} {\bibfnamefont
  {M.}~\bibnamefont {M{\ifmmode\ddot{u}\else\"{u}\fi}ller}}, \bibinfo {author}
  {\bibfnamefont {D.}~\bibnamefont {Neumaier}}, \bibinfo {author}
  {\bibfnamefont {K.}~\bibnamefont {Watanabe}}, \bibinfo {author}
  {\bibfnamefont {T.}~\bibnamefont {Taniguchi}}, \bibinfo {author}
  {\bibfnamefont {F.}~\bibnamefont {Libisch}}, \bibinfo {author} {\bibfnamefont
  {B.}~\bibnamefont {Beschoten}}, \bibinfo {author} {\bibfnamefont
  {F.}~\bibnamefont {Hassler}}, \ and\ \bibinfo {author} {\bibfnamefont
  {C.}~\bibnamefont {Stampfer}},\ }\href {\doibase
  10.1103/PhysRevLett.124.177701} {\bibfield  {journal} {\bibinfo  {journal}
  {Phys. Rev. Lett.}\ }\textbf {\bibinfo {volume} {124}},\ \bibinfo {pages}
  {177701} (\bibinfo {year} {2020}{\natexlab{c}})}\BibitemShut {NoStop}%
\bibitem [{\citenamefont {Bonet}\ \emph {et~al.}(2002)\citenamefont {Bonet},
  \citenamefont {Deshmukh},\ and\ \citenamefont {Ralph}}]{Bonet2002Jan}%
  \BibitemOpen
  \bibfield  {author} {\bibinfo {author} {\bibfnamefont {E.}~\bibnamefont
  {Bonet}}, \bibinfo {author} {\bibfnamefont {M.~M.}\ \bibnamefont {Deshmukh}},
  \ and\ \bibinfo {author} {\bibfnamefont {D.~C.}\ \bibnamefont {Ralph}},\
  }\href {\doibase 10.1103/PhysRevB.65.045317} {\bibfield  {journal} {\bibinfo
  {journal} {Phys. Rev. B}\ }\textbf {\bibinfo {volume} {65}},\ \bibinfo
  {pages} {045317} (\bibinfo {year} {2002})}\BibitemShut {NoStop}%
\bibitem [{\citenamefont {Knothe}\ \emph {et~al.}(2022)\citenamefont {Knothe},
  \citenamefont {Glazman},\ and\ \citenamefont {Fal{'}ko}}]{Knothe2022Apr}%
  \BibitemOpen
  \bibfield  {author} {\bibinfo {author} {\bibfnamefont {A.}~\bibnamefont
  {Knothe}}, \bibinfo {author} {\bibfnamefont {L.~I.}\ \bibnamefont {Glazman}},
  \ and\ \bibinfo {author} {\bibfnamefont {V.~I.}\ \bibnamefont {Fal{'}ko}},\
  }\href {\doibase 10.1088/1367-2630/ac5d00} {\bibfield  {journal} {\bibinfo
  {journal} {New J. Phys.}\ }\textbf {\bibinfo {volume} {24}},\ \bibinfo
  {pages} {043003} (\bibinfo {year} {2022})}\BibitemShut {NoStop}%
\bibitem [{\citenamefont {M{\ifmmode\ddot{o}\else\"{o}\fi}ller}\ \emph
  {et~al.}(2021)\citenamefont {M{\ifmmode\ddot{o}\else\"{o}\fi}ller},
  \citenamefont {Banszerus}, \citenamefont {Knothe}, \citenamefont {Steiner},
  \citenamefont {Icking}, \citenamefont {Trellenkamp}, \citenamefont {Lentz},
  \citenamefont {Watanabe}, \citenamefont {Taniguchi}, \citenamefont {Glazman},
  \citenamefont {Fal{'}ko}, \citenamefont {Volk},\ and\ \citenamefont
  {Stampfer}}]{Moller2021Dec}%
  \BibitemOpen
  \bibfield  {author} {\bibinfo {author} {\bibfnamefont {S.}~\bibnamefont
  {M{\ifmmode\ddot{o}\else\"{o}\fi}ller}}, \bibinfo {author} {\bibfnamefont
  {L.}~\bibnamefont {Banszerus}}, \bibinfo {author} {\bibfnamefont
  {A.}~\bibnamefont {Knothe}}, \bibinfo {author} {\bibfnamefont
  {C.}~\bibnamefont {Steiner}}, \bibinfo {author} {\bibfnamefont
  {E.}~\bibnamefont {Icking}}, \bibinfo {author} {\bibfnamefont
  {S.}~\bibnamefont {Trellenkamp}}, \bibinfo {author} {\bibfnamefont
  {F.}~\bibnamefont {Lentz}}, \bibinfo {author} {\bibfnamefont
  {K.}~\bibnamefont {Watanabe}}, \bibinfo {author} {\bibfnamefont
  {T.}~\bibnamefont {Taniguchi}}, \bibinfo {author} {\bibfnamefont {L.~I.}\
  \bibnamefont {Glazman}}, \bibinfo {author} {\bibfnamefont {V.~I.}\
  \bibnamefont {Fal{'}ko}}, \bibinfo {author} {\bibfnamefont {C.}~\bibnamefont
  {Volk}}, \ and\ \bibinfo {author} {\bibfnamefont {C.}~\bibnamefont
  {Stampfer}},\ }\href {\doibase 10.1103/PhysRevLett.127.256802} {\bibfield
  {journal} {\bibinfo  {journal} {Phys. Rev. Lett.}\ }\textbf {\bibinfo
  {volume} {127}},\ \bibinfo {pages} {256802} (\bibinfo {year}
  {2021})}\BibitemShut {NoStop}%
\bibitem [{\citenamefont {Mani}\ \emph {et~al.}(2012)\citenamefont {Mani},
  \citenamefont {Hankinson}, \citenamefont {Berger},\ and\ \citenamefont
  {de~Heer}}]{Mani2012Aug}%
  \BibitemOpen
  \bibfield  {author} {\bibinfo {author} {\bibfnamefont {R.~G.}\ \bibnamefont
  {Mani}}, \bibinfo {author} {\bibfnamefont {J.}~\bibnamefont {Hankinson}},
  \bibinfo {author} {\bibfnamefont {C.}~\bibnamefont {Berger}}, \ and\ \bibinfo
  {author} {\bibfnamefont {W.~A.}\ \bibnamefont {de~Heer}},\ }\href {\doibase
  10.1038/ncomms1986} {\bibfield  {journal} {\bibinfo  {journal} {Nat.
  Commun.}\ }\textbf {\bibinfo {volume} {3}},\ \bibinfo {pages} {996} (\bibinfo
  {year} {2012})}\BibitemShut {NoStop}%
\bibitem [{\citenamefont {Sichau}\ \emph {et~al.}(2019)\citenamefont {Sichau},
  \citenamefont {Prada}, \citenamefont {Anlauf}, \citenamefont {Lyon},
  \citenamefont {Bosnjak}, \citenamefont {Tiemann},\ and\ \citenamefont
  {Blick}}]{Sichau2019Feb}%
  \BibitemOpen
  \bibfield  {author} {\bibinfo {author} {\bibfnamefont {J.}~\bibnamefont
  {Sichau}}, \bibinfo {author} {\bibfnamefont {M.}~\bibnamefont {Prada}},
  \bibinfo {author} {\bibfnamefont {T.}~\bibnamefont {Anlauf}}, \bibinfo
  {author} {\bibfnamefont {T.~J.}\ \bibnamefont {Lyon}}, \bibinfo {author}
  {\bibfnamefont {B.}~\bibnamefont {Bosnjak}}, \bibinfo {author} {\bibfnamefont
  {L.}~\bibnamefont {Tiemann}}, \ and\ \bibinfo {author} {\bibfnamefont
  {R.~H.}\ \bibnamefont {Blick}},\ }\href {\doibase
  10.1103/PhysRevLett.122.046403} {\bibfield  {journal} {\bibinfo  {journal}
  {Phys. Rev. Lett.}\ }\textbf {\bibinfo {volume} {122}},\ \bibinfo {pages}
  {046403} (\bibinfo {year} {2019})}\BibitemShut {NoStop}%
\bibitem [{\citenamefont {Lyon}\ \emph {et~al.}(2017)\citenamefont {Lyon},
  \citenamefont {Sichau}, \citenamefont {Dorn}, \citenamefont {Centeno},
  \citenamefont {Pesquera}, \citenamefont {Zurutuza},\ and\ \citenamefont
  {Blick}}]{Lyon2017Aug}%
  \BibitemOpen
  \bibfield  {author} {\bibinfo {author} {\bibfnamefont {T.~J.}\ \bibnamefont
  {Lyon}}, \bibinfo {author} {\bibfnamefont {J.}~\bibnamefont {Sichau}},
  \bibinfo {author} {\bibfnamefont {A.}~\bibnamefont {Dorn}}, \bibinfo {author}
  {\bibfnamefont {A.}~\bibnamefont {Centeno}}, \bibinfo {author} {\bibfnamefont
  {A.}~\bibnamefont {Pesquera}}, \bibinfo {author} {\bibfnamefont
  {A.}~\bibnamefont {Zurutuza}}, \ and\ \bibinfo {author} {\bibfnamefont
  {R.~H.}\ \bibnamefont {Blick}},\ }\href {\doibase
  10.1103/PhysRevLett.119.066802} {\bibfield  {journal} {\bibinfo  {journal}
  {Phys. Rev. Lett.}\ }\textbf {\bibinfo {volume} {119}},\ \bibinfo {pages}
  {066802} (\bibinfo {year} {2017})}\BibitemShut {NoStop}%
\bibitem [{\citenamefont {Johnson}\ \emph {et~al.}(2005)\citenamefont
  {Johnson}, \citenamefont {Petta}, \citenamefont {Marcus}, \citenamefont
  {Hanson},\ and\ \citenamefont {Gossard}}]{Johnson2005Oct}%
  \BibitemOpen
  \bibfield  {author} {\bibinfo {author} {\bibfnamefont {A.~C.}\ \bibnamefont
  {Johnson}}, \bibinfo {author} {\bibfnamefont {J.~R.}\ \bibnamefont {Petta}},
  \bibinfo {author} {\bibfnamefont {C.~M.}\ \bibnamefont {Marcus}}, \bibinfo
  {author} {\bibfnamefont {M.~P.}\ \bibnamefont {Hanson}}, \ and\ \bibinfo
  {author} {\bibfnamefont {A.~C.}\ \bibnamefont {Gossard}},\ }\href {\doibase
  10.1103/PhysRevB.72.165308} {\bibfield  {journal} {\bibinfo  {journal} {Phys.
  Rev. B}\ }\textbf {\bibinfo {volume} {72}},\ \bibinfo {pages} {165308}
  (\bibinfo {year} {2005})}\BibitemShut {NoStop}%
\bibitem [{\citenamefont {Borselli}\ \emph {et~al.}(2011)\citenamefont
  {Borselli}, \citenamefont {Eng}, \citenamefont {Croke}, \citenamefont
  {Maune}, \citenamefont {Huang}, \citenamefont {Ross}, \citenamefont
  {Kiselev}, \citenamefont {Deelman}, \citenamefont {Alvarado-Rodriguez},
  \citenamefont {Schmitz}, \citenamefont {Sokolich}, \citenamefont {Holabird},
  \citenamefont {Hazard}, \citenamefont {Gyure},\ and\ \citenamefont
  {Hunter}}]{Borselli2011Aug}%
  \BibitemOpen
  \bibfield  {author} {\bibinfo {author} {\bibfnamefont {M.~G.}\ \bibnamefont
  {Borselli}}, \bibinfo {author} {\bibfnamefont {K.}~\bibnamefont {Eng}},
  \bibinfo {author} {\bibfnamefont {E.~T.}\ \bibnamefont {Croke}}, \bibinfo
  {author} {\bibfnamefont {B.~M.}\ \bibnamefont {Maune}}, \bibinfo {author}
  {\bibfnamefont {B.}~\bibnamefont {Huang}}, \bibinfo {author} {\bibfnamefont
  {R.~S.}\ \bibnamefont {Ross}}, \bibinfo {author} {\bibfnamefont {A.~A.}\
  \bibnamefont {Kiselev}}, \bibinfo {author} {\bibfnamefont {P.~W.}\
  \bibnamefont {Deelman}}, \bibinfo {author} {\bibfnamefont {I.}~\bibnamefont
  {Alvarado-Rodriguez}}, \bibinfo {author} {\bibfnamefont {A.~E.}\ \bibnamefont
  {Schmitz}}, \bibinfo {author} {\bibfnamefont {M.}~\bibnamefont {Sokolich}},
  \bibinfo {author} {\bibfnamefont {K.~S.}\ \bibnamefont {Holabird}}, \bibinfo
  {author} {\bibfnamefont {T.~M.}\ \bibnamefont {Hazard}}, \bibinfo {author}
  {\bibfnamefont {M.~F.}\ \bibnamefont {Gyure}}, \ and\ \bibinfo {author}
  {\bibfnamefont {A.~T.}\ \bibnamefont {Hunter}},\ }\href {\doibase
  10.1063/1.3623479} {\bibfield  {journal} {\bibinfo  {journal} {Appl. Phys.
  Lett.}\ }\textbf {\bibinfo {volume} {99}},\ \bibinfo {pages} {063109}
  (\bibinfo {year} {2011})}\BibitemShut {NoStop}%
\bibitem [{\citenamefont {Tong}\ \emph {et~al.}(2022)\citenamefont {Tong},
  \citenamefont {Kurzmann}, \citenamefont {Garreis}, \citenamefont {Huang},
  \citenamefont {Jele}, \citenamefont {Eich}, \citenamefont {Ginzburg},
  \citenamefont {Mittag}, \citenamefont {Watanabe}, \citenamefont {Taniguchi},
  \citenamefont {Ensslin},\ and\ \citenamefont {Ihn}}]{Tong2022Feb}%
  \BibitemOpen
  \bibfield  {author} {\bibinfo {author} {\bibfnamefont {C.}~\bibnamefont
  {Tong}}, \bibinfo {author} {\bibfnamefont {A.}~\bibnamefont {Kurzmann}},
  \bibinfo {author} {\bibfnamefont {R.}~\bibnamefont {Garreis}}, \bibinfo
  {author} {\bibfnamefont {W.~W.}\ \bibnamefont {Huang}}, \bibinfo {author}
  {\bibfnamefont {S.}~\bibnamefont {Jele}}, \bibinfo {author} {\bibfnamefont
  {M.}~\bibnamefont {Eich}}, \bibinfo {author} {\bibfnamefont {L.}~\bibnamefont
  {Ginzburg}}, \bibinfo {author} {\bibfnamefont {C.}~\bibnamefont {Mittag}},
  \bibinfo {author} {\bibfnamefont {K.}~\bibnamefont {Watanabe}}, \bibinfo
  {author} {\bibfnamefont {T.}~\bibnamefont {Taniguchi}}, \bibinfo {author}
  {\bibfnamefont {K.}~\bibnamefont {Ensslin}}, \ and\ \bibinfo {author}
  {\bibfnamefont {T.}~\bibnamefont {Ihn}},\ }\href {\doibase
  10.1103/PhysRevLett.128.067702} {\bibfield  {journal} {\bibinfo  {journal}
  {Phys. Rev. Lett.}\ }\textbf {\bibinfo {volume} {128}},\ \bibinfo {pages}
  {067702} (\bibinfo {year} {2022})}\BibitemShut {NoStop}%
\bibitem [{\citenamefont {Albrecht}\ \emph {et~al.}(2017)\citenamefont
  {Albrecht}, \citenamefont {Moers},\ and\ \citenamefont
  {Hermanns}}]{Albrecht2017May}%
  \BibitemOpen
  \bibfield  {author} {\bibinfo {author} {\bibfnamefont {W.}~\bibnamefont
  {Albrecht}}, \bibinfo {author} {\bibfnamefont {J.}~\bibnamefont {Moers}}, \
  and\ \bibinfo {author} {\bibfnamefont {B.}~\bibnamefont {Hermanns}},\ }\href
  {\doibase 10.17815/jlsrf-3-158} {\bibfield  {journal} {\bibinfo  {journal}
  {Journal of Large-Scale Research Facilities}\ }\textbf {\bibinfo {volume}
  {3}},\ \bibinfo {pages} {112} (\bibinfo {year} {2017})}\BibitemShut {NoStop}%
\end{thebibliography}

%

\end{document}


\author{L.~Banszerus}
\thanks{These two authors contributed equally.}
\author{S.~M\"oller}
\thanks{These two authors contributed equally.}
\affiliation{JARA-FIT and 2nd Institute of Physics, RWTH Aachen University, 52074 Aachen, Germany,~EU}%
\affiliation{Peter Gr\"unberg Institute  (PGI-9), Forschungszentrum J\"ulich, 52425 J\"ulich,~Germany,~EU}
\author{K.~Hecker}
\author{E.~Icking}
\affiliation{JARA-FIT and 2nd Institute of Physics, RWTH Aachen University, 52074 Aachen, Germany,~EU}%
\affiliation{Peter Gr\"unberg Institute  (PGI-9), Forschungszentrum J\"ulich, 52425 J\"ulich,~Germany,~EU}
\author{K.~Watanabe}
\affiliation{Research Center for Functional Materials, 
National Institute for Materials Science, 1-1 Namiki, Tsukuba 305-0044, Japan}
\author{T.~Taniguchi}
\affiliation{International Center for Materials Nanoarchitectonics, 
National Institute for Materials Science,  1-1 Namiki, Tsukuba 305-0044, Japan}%
\author{F.~Hassler}
\affiliation{JARA-Institute for Quantum Information, RWTH Aachen University, 52056 Aachen, Germany, EU}
\author{C.~Volk}
\author{C.~Stampfer}
\email{stampfer@physik.rwth-aachen.de}
\affiliation{JARA-FIT and 2nd Institute of Physics, RWTH Aachen University, 52074 Aachen, Germany,~EU}%
\affiliation{Peter Gr\"unberg Institute  (PGI-9), Forschungszentrum J\"ulich, 52425 J\"ulich,~Germany,~EU}%

\title{Supporting Information:\\ Particle-hole symmetry protects spin-valley blockade in graphene quantum dots}

\date{\today}

\maketitle
\tableofcontents

\clearpage
\subsection{Charge stability diagrams for opposite bias voltages in DQD \#1}

\begin{figure}[!thb]
\centering
\includegraphics[draft=false,keepaspectratio=true,clip,width=\linewidth]{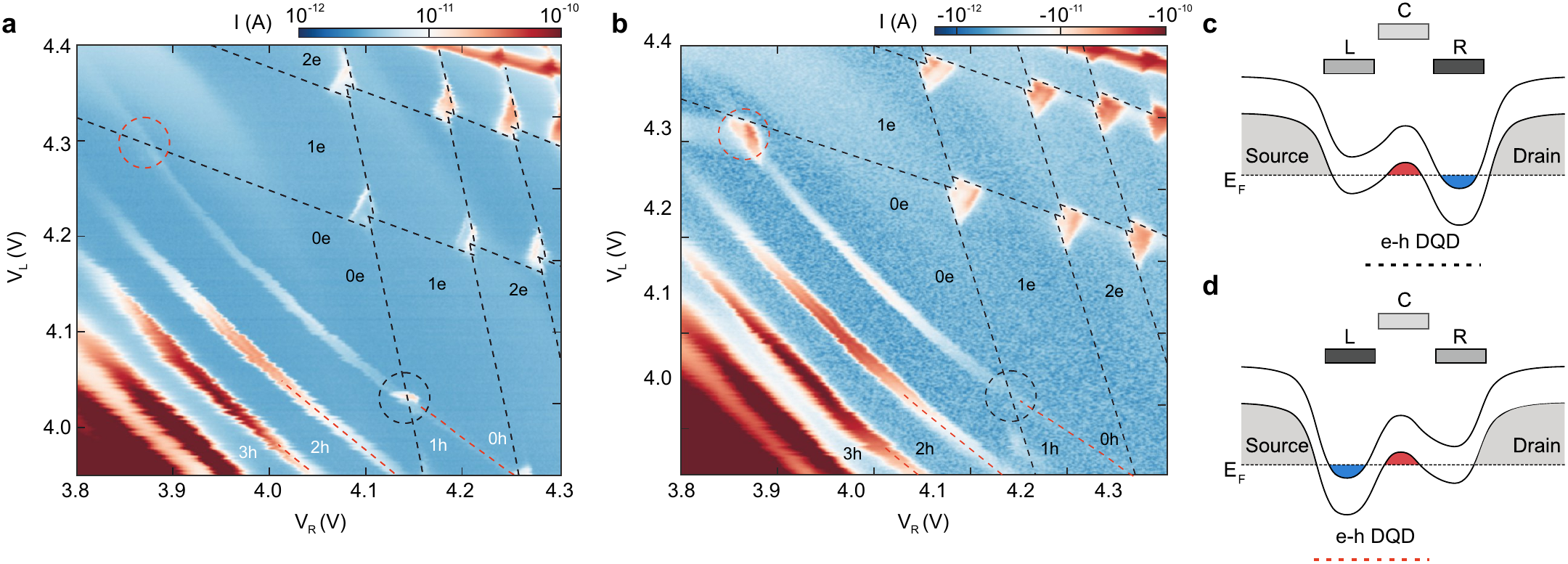}
\caption[Fig01]{Charge stability diagrams of DQD \#1 (as in Fig.~1d of the main text) measured at a bias voltage of \textbf{a} $V_\mathrm{SD} = 1$~mV and \textbf{b} $V_\mathrm{SD} = -1$mV (T=10mK).
The dashed circles mark the formation of single electron -- single hole DQDs using the hole QD and an electron QD to the left (red) or right (black) of the hole QD. \textbf{c-d} Schematics of the valence and conduction band edge profiles along the p-type
channel. An electron-hole double quantum dot is formed using the hole QD and the electron QD underneath the left (right) FG (see red (black) circles in Fig.~S1a,b).
 }
\label{S1}
\end{figure}

Fig.~\ref{S1} compares charge stability diagrams measured at positive and negative bias voltage in DQD \#1 (c.f. Figs. 1, 2 and 3 in the main text). 
The dashed lines indicate the charge transitions of the electron (black) and hole (red) QDs. 
Electron-hole (e-h) DQDs are formed at the intersections of these charging lines. 
For the left electron-hole DQD ($(0h,0e)\leftrightarrow(1h,1e)$ transition, see red circle), transport is blocked at positive bias, while for the right electron-hole DQD ($(0h,0e)\leftrightarrow(1h,1e)$ transition, see black circle), transport is blocked at negative bias. The data in the main text has been obtained in the latter regime.

\clearpage
\subsection{Extracting $\Delta_\text{SO}$ from measurements on a single-electron DQD in the same device}
\begin{figure}[!thb]
\centering
\includegraphics[draft=false,keepaspectratio=true,clip,width=0.9\linewidth]{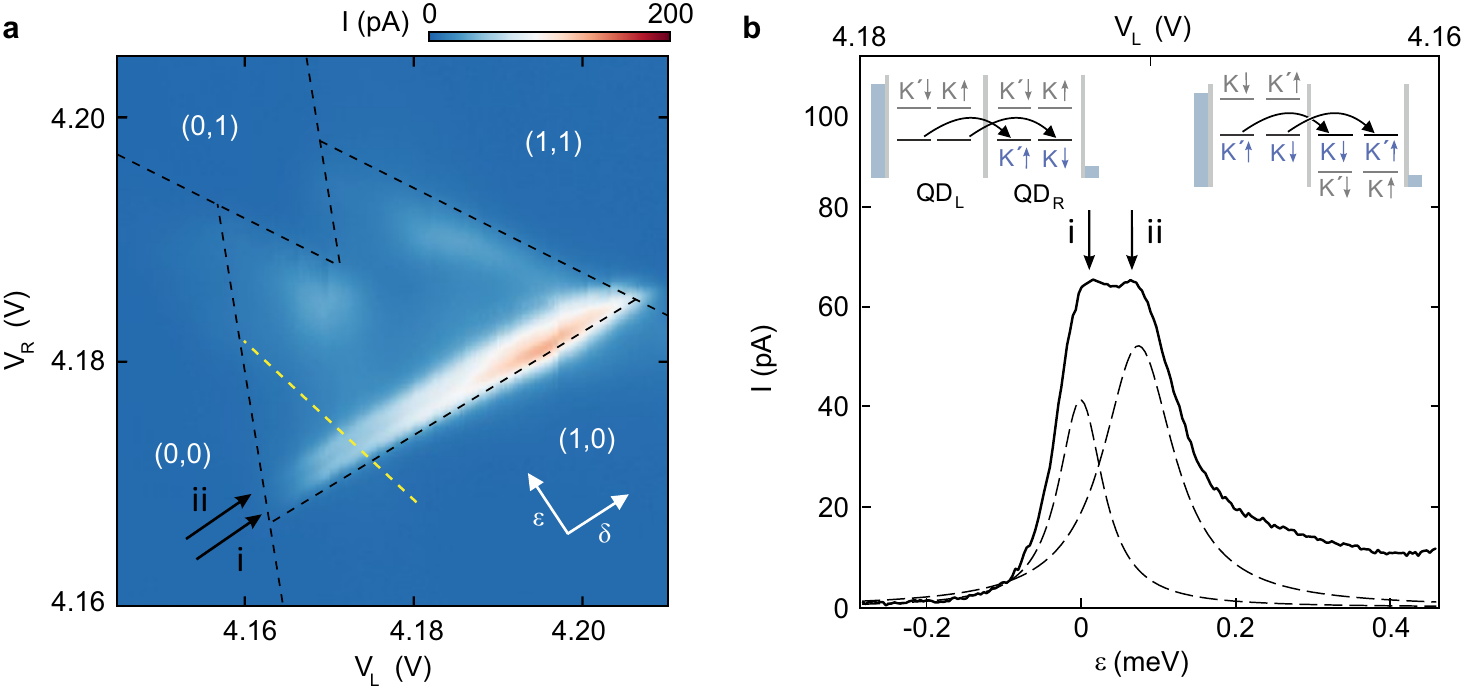}
\caption[Fig01]{
\textbf{a} Charge stability diagrams of the $(1e,0e)\leftrightarrow(0e,1e)$ transition of an electron-electron DQD measured at $V_\mathrm{SD} = 1~$mV and $B_\perp = 0~$T (T=10mK).
A ground state and an excited state transition are visible (see black arrows). 
\textbf{b} Cut along the yellow dashed line in a. Two Lorentzian peaks (dashed lines) are fitted to the data.
Inset: Schematic energy diagrams of an electron-electron DQD in the finite bias regime for different interdot detuning
energies $\varepsilon$, illustrating resonant transport from the left (L) to the right (R) QD through the ground state of each QD (transition (i)) and resonant transport at $\varepsilon = \Delta_\mathrm{SO}$ (transition (ii)).
}
\label{S5}
\end{figure}

To compare the measured value for $\Delta_\mathrm{SO}$ in the electron-hole DQD and to demonstrate that the magnitude of the SO gap is symmetric for electrons and holes, we present measurements of  $\Delta_\mathrm{SO}$ in an electron-electron DQD. Fig.~\ref{S5} shows a close-up of the first triple point of an electron-electron DQD formed in the same device (c.f. Fig.~\ref{S1}). Transport via a ground state and an excited state can be observed. 
We extract their energy splitting by fitting two Lorentzian peaks to a linecut through the triple point (see Fig.~\ref{S5}b). The determined value of $\Delta_\mathrm{SO} = 68 \pm 7~\mu$eV is in good agreement with the ones observed in the electron-hole DQD regime. A detailed discussion of $\Delta_\mathrm{SO}$ and the single particle spectrum in the electron DQD in this device is given in Ref.~\cite{Banszerus2021Sep}.

\clearpage
\subsection{Additional data set for another e-h double quantum dot (DQD \#2) in the same device }

\begin{figure}[!thb]
\centering
\includegraphics[draft=false,keepaspectratio=true,clip,width=0.85\linewidth]{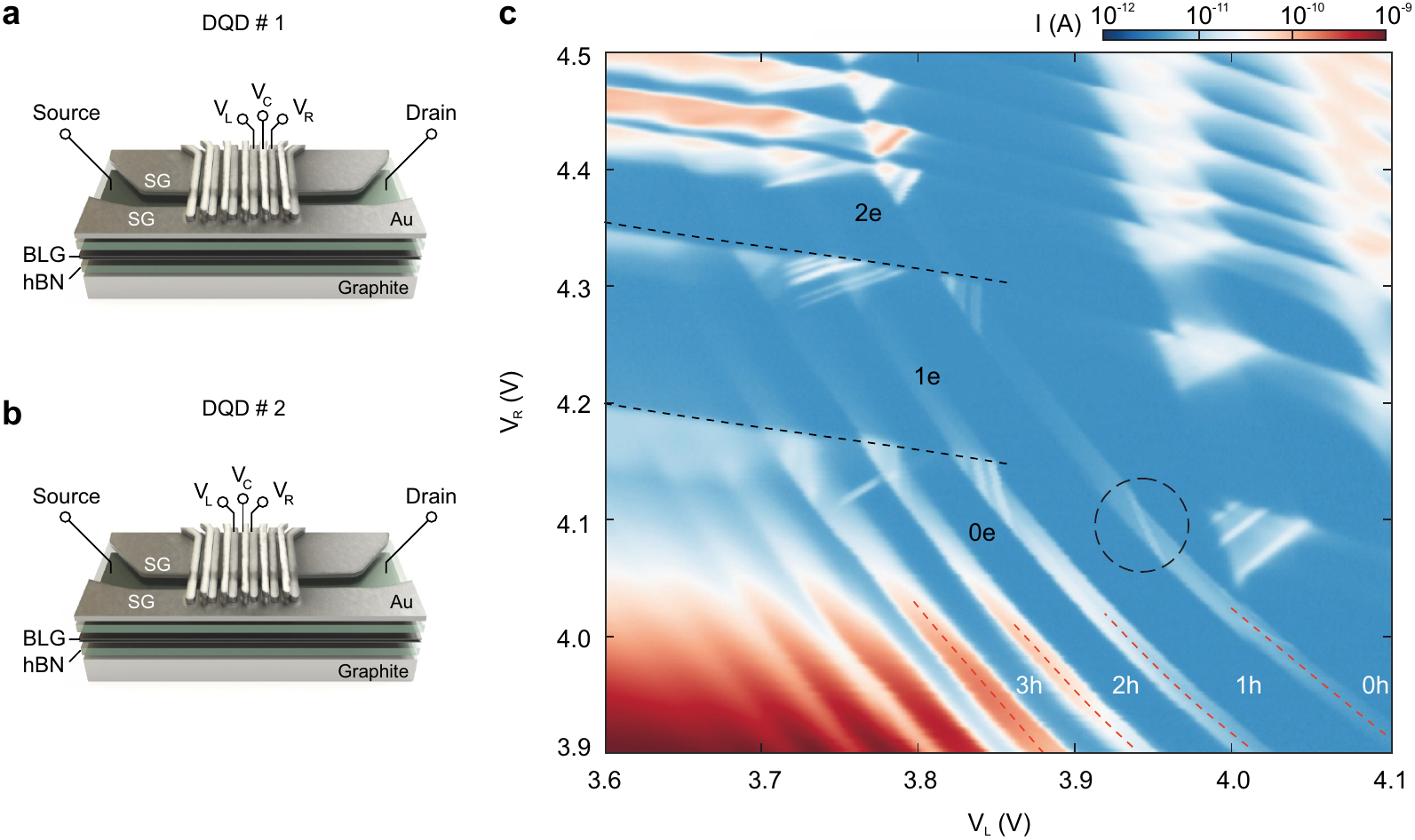}
\caption[Fig03]{ \textbf{a} and \textbf{b} Gate configurations used to form DQD \#1 and DQD \#2 in the device, respectively. \textbf{c}
 Charge stability diagram of an e-h DQD formed with the second set of gate fingers (DQD \#2, see panel b). The dashed circle marks the $(0h,0e)\rightarrow(1h,1e)$ transition. $V_\mathrm{SD} = 1~$mV (T=10mK).
}
\label{S3}
\end{figure}

\begin{figure}[!thb]
\centering
\includegraphics[draft=false,keepaspectratio=true,clip,width=1\linewidth]{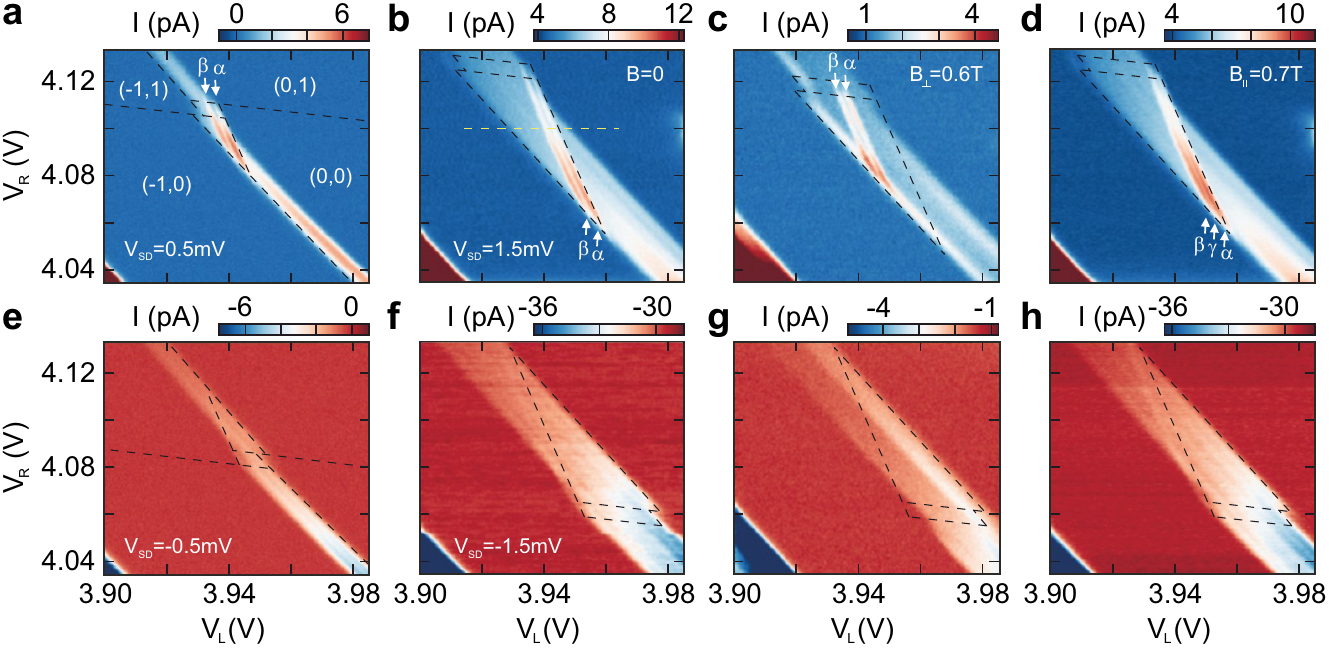}
\caption[Fig03]{  
\textbf{a, b} Close-ups of the $(0h,0e)\rightarrow(1h,1e)$ triple point at $V_\mathrm{SD} = 0.5~$mV and $V_\mathrm{SD} = 1.5~$mV, respectively. Transport only occurs via the $\alpha$ and $\beta$ transition.
\textbf{c} Charge stability diagram as in c measured at $B_\perp = 0.6$~T. 
\textbf{d} Charge stability diagram as in b at $B_\parallel = 0.7$~T.
\textbf{e, f} Charge stability diagrams as in b and c at $V_\mathrm{SD} = -0.5$~mV and $V_\mathrm{SD} = -1.5$~mV. Transport is strongly suppressed, only co-tunneling can be observed. 
\textbf{g, h}  Charge stability diagrams as in g measured at $B_\perp = 0.6$~T and $B_\parallel = 0.7$~T.
}
\label{S32}
\end{figure}

A second e-h DQD has been studied, formed with a different set of gate fingers on the same gated bilayer graphene device as presented in the main text (DQD \#2 depicted in Fig.~S2b). 
The single electron -- single hole transition, $(0h,0e)\rightarrow(1h,1e)$, is highlighted by the dashed circle in the charge stability diagram (see Fig.~\ref{S3}c). 

Measurements of that bias triangle are shown in Figs.~\ref{S32} for different $V_\mathrm{SD}$ and magnetic fields, showing good agreement with the data presented for DQD \#1 in Fig.~2.
In contrast to the data presented in the main manuscript, co-tunneling is more pronounced due to a strong coupling of the hole QD to the reservoir.

\begin{figure}[!thb]
\centering
\includegraphics[draft=false,keepaspectratio=true,clip,width=\linewidth]{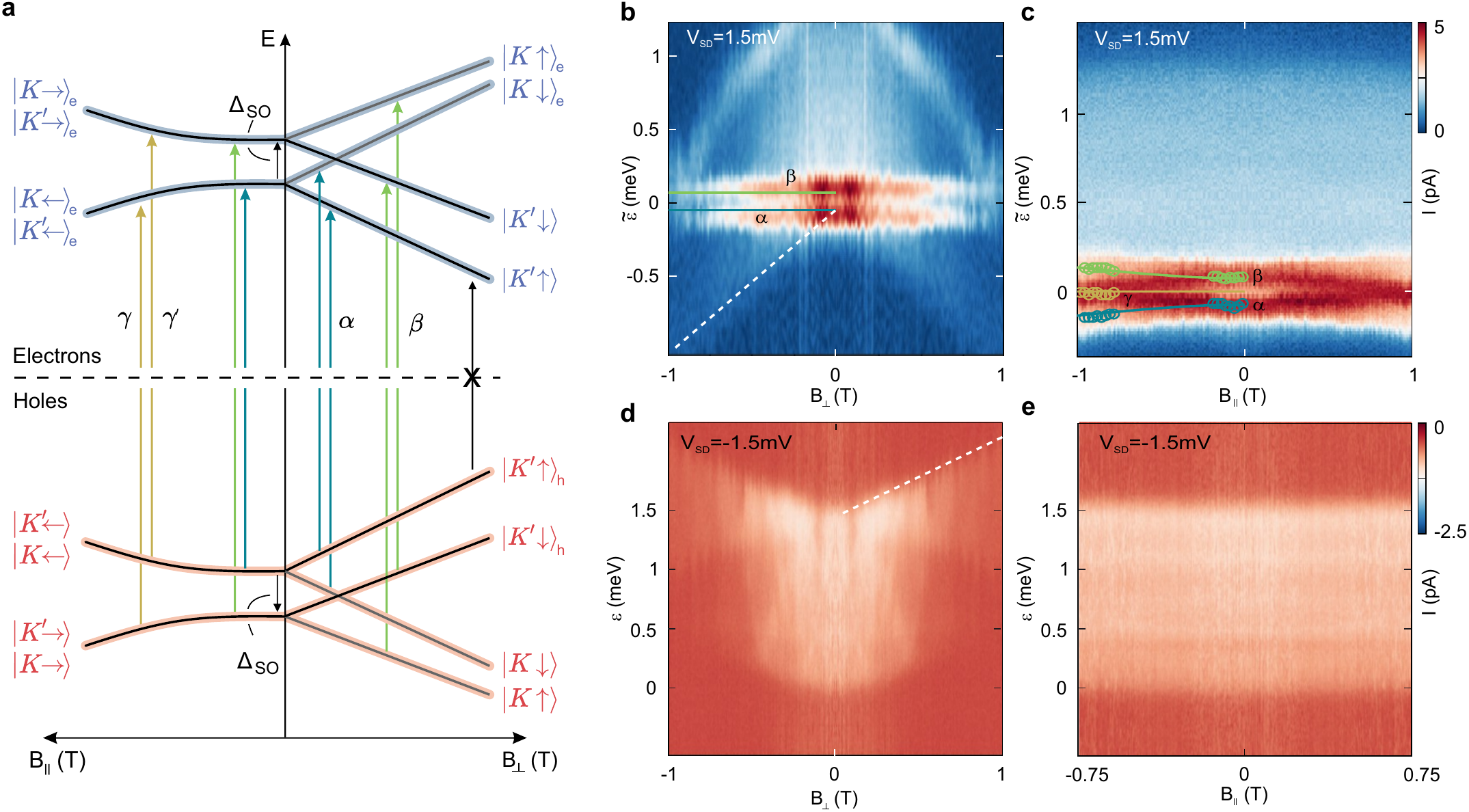}
\caption[Fig01]{
\textbf{a} Energy dispersion of single-particle states in the first orbital for electrons and holes as a function of in-plane ($B_\parallel$, left) and out-of-plane ($B_\perp$, right) magnetic fields. States and transitions are labelled as in Fig.~3a of the main text.
\textbf{b} Current through DQD \# 2 as a function of the detuning energy $\widetilde \varepsilon$ (see yellow dashed line in Fig.~\ref{S32}b) and $B_\perp$ at $V_\mathrm{SD} = 1.5~$mV. The white dashed line marks the onset of the bias transport window.
\textbf{c} Current through the device as a function of $\widetilde \varepsilon$ and $B_\parallel$ at $V_\mathrm{SD} = 1.5~$mV. \textbf{d, e} Data acquired in the blockade regime ($V_\mathrm{SD} = -1.5~$mV). The current has been measured as a function of $B_\perp$ and $B_\parallel$, respectively. Data has been symmetrized around $B = 0$.
}
\label{S4}
\end{figure}

The magnetic field dependent spectrum of the first electron and the first hole states is depicted in Fig.~\ref{S4}a (c.f. Fig.~3a of the main text).
Figs.~\ref{S4}b and c show measurements complementary to the one presented in Fig.~3 of the main text, recorded for DQD \#2 shown in Fig.~\ref{S3}b. 
The measurements show that the difference in detuning energy between  $\alpha$ and $\beta$ is independent of $B_\perp$, the energy splitting measures $\Delta \varepsilon = 150 \pm 10 \,\mu$eV, which corresponds to $2\Delta_\mathrm{SO}$.
The background current originates from co-tunneling in the bias transport window (its onset is highlighted by the white dashed line), which shifts in energy with increasing $|B_\perp|$. This is due to the fact that the bias window is defined by the (forbidden) ground state transition $\ket{K'\uparrow}_e \leftrightarrow  \ket{K'\uparrow}_h$, which requires less detuning for increasing $|B_\perp|$.
The same measurement for parallel magnetic fields shows the effect of the spins being continuously canted into the BLG plane. 
The difference in detuning of the transitions $\alpha$ and $\beta$ increases while a third resonance, $\gamma$, emerges.
The data is in good qualitative and quantitative agreement with the data presented in Fig.~3 of the main text. 
Figs.~\ref{S4}d and \ref{S4}e show magneto-transport data in the single-particle blockade regime. The spin-valley blockade is not lifted under the influence of both in-plane and out-of-plane magnetic fields. 
Transport via co-tunneling is suppressed at increasing $B_\perp$ as (also in this case) the tunneling barriers turn more opaque due to magnetic confinement.

\clearpage
\subsection{Simulation of magnetotransport through an e-h DQD}
We simulate transport within the DQD bias triangles and along the detuning cuts by solving the rate equations for the electron and hole QD states presented in Fig.~3a following the approach used in Ref.~\cite{Knothe2022Apr}. The energy of the respective electron and hole states is given by

\begin{align}
    \mathrm{H_{e}} &= \frac{1}{2} \Delta_\text{SO} \tau_z s_z + \frac{1}{2} g_\text{s} \mu_\text{B} \mathbf{B} \cdot \mathbf{s} + \frac{1}{2} g_\text{v} \mu_\text{B} \mathrm{B}_z \tau_z  \\
    \mathrm{H_h} &=  - \mathrm{H_{e}},
\end{align}
with the spin and valley g-factors $g_\text{s} = 2 $ and $g_\text{v} = 15$,
the Bohr magneton $\mu_\text{B}$, the proximity enhanced (intrinsic) Kane-Mele spin-orbit coupling $\Delta_\text{SO} = 70\, \mu$eV and the Pauli matrices $s_i$ and $\tau_i$ which act on spin and valley, respectively. We approximate the effect of the right (R) and left (L) finger gate on the charging energy of the system by 
\begin{equation}
    E_\text{c} (N_\text{R}, N_\text{L}) = e N_\text{R}  V_\text{R} + e N_\text{L}  V_\text{L} \,,
\end{equation}
with the absolute value of the elementary charge, $e$, the QD occupation number $N_\text{L} = -1$ (1h),  $N_\text{R} = 1$ (1e) and the gate voltages $V_\text{R}$ and $V_\text{L}$. For describing transport through the e-h DQD, we focus on the $(0,0) \rightarrow (-1,1) \rightarrow (-1, 0) \rightarrow (0,0)$ charge cycle and only consider sequential tunneling. There are in total 25 possible states of the system $\chi$ = (hole QD state, electron QD state) with
\begin{align}
    \chi &= (\overline{\phi_\text{h}}\, , \overline{\psi_\text{e}}) \\
    \overline{\phi_\text{h}}\, , \overline{\psi_\text{e}}  &\in \{0,K{\uparrow},K{\downarrow},K'{\uparrow},K'{\downarrow} \}.  
\end{align}
Here, $ \overline{\phi_\text{h}}\, , \overline{\psi_\text{e}}$ describe the state of the left and right QD, which includes the four single particle states, as well as the QD being empty.

We assume no mixing between lead and QD states and equal tunnel probabilities to and from the leads for all states, $\gamma^\text{L,R} = 1.7$ GHz. Thus, we obtain the transition rates between QD states involving tunneling processes from the leads (L,R) by computing
\begin{align}
W^\text{L,R}_{\chi \leftarrow \chi^{\prime}} &= \gamma^\text{L,R} \,  f(E_{\chi} - E_{\chi^{\prime}} - \mu^\text{L,R}),
\end{align}
with the Fermi-function, $f$, at $T = 0.1\,$K, and the electron and hole QD states $\phi_\text{h}, \psi_\text{e}$. Note that hole states only tunnel to the left lead and electron states only tunnel to the right lead. 

For interdot transitions, we assume no mixing of electron and hole states due to the small interdot tunnel coupling. For simplicity, relaxation is neglected. We obtain the rates of the interdot transition by computing 
\begin{align}
W^\text{inter}_{(0,0) \leftarrow (\phi_\text{h}, \psi_\text{e})} &=  W^\text{inter}_{(\phi_\text{h}, \psi_\text{e}) \leftarrow (0,0) }= G^\text{inter} \, \braket{\phi_\text{h} | \psi_\text{e}}  \frac{1}{\sqrt{2 \pi \sigma}} \mathrm{exp}\left(  {-\frac{(E_{(0,0)} - E_{(\phi_\text{h}, \psi_\text{e})})^2}{4 \sigma^2}} \right),
\label{align:interdot}
\end{align} 
 with the interdot tunnel rate $\gamma^\text{inter} = 6$ kHz and $\phi_\text{h}, \psi_\text{e} \in \{ K{\uparrow},K{\downarrow},K'{\uparrow},K'{\downarrow} \}$. The Gaussian energy smearing models the experimentally observed peaks with an estimated width of the resonances $\Gamma = 40 \,\mu$eV. We expect that this smearing originates from voltage fluctuations of the finger gates. The overlap between electron and hole states is given by $\braket{\phi_\text{h} | \psi_\text{e}} = (\sigma_y \tau_x s_y)_{\phi_\text{h}, \psi_\text{e}}$ in order to assure that only electrons and holes with opposite quantum numbers are created (or annihilated). With equation (\ref{align:interdot}) we implicitly assume that the states in the left and right QD have no coherent phase relation. 

We solve the master equation of the probabilities, $\mathrm{P}_{\chi}$, for the system to be in state $\chi$,
\begin{equation}
\dot{\mathrm{P}}_{\chi} =\sum_{\chi'} (W_{ \chi \leftarrow \chi'} \, \dot{\mathrm{P}}_{\chi'}  - W_{ \chi' \leftarrow \chi} \, \dot{\mathrm{P}}_{\chi} ) ,
\label{eqn:rateeqn}
\end{equation}

in the stationary limit, $\dot{\mathrm{P}}_{\chi} = 0$, normalizing the probabilities to $\sum_{\chi} \mathrm{P}_{\chi} = 1$.
In the stationary limit, we can compute the current through the double QD by computing the current flow from the right QD to lead R:
\begin{equation} 
 I^\text{R}= e \sum_{\phi_\text{h}, \psi_\text{e} } \left( W^\text{R}_{(\phi_\text{h}, 0) \leftarrow (\phi_\text{h}, \psi_\text{e})} \dot{\mathrm{P}}_{(\phi_\text{h}, \psi_\text{e})}
-  W^\text{R}_{(\phi_\text{h}, \psi_\text{e}) \leftarrow (\phi_\text{h}, 0) } \dot{\mathrm{P}}_{(\phi_\text{h}, 0)} \right).
 \label{eqn:Iseq}
\end{equation}

\begin{figure}[!thb]
\centering
\includegraphics[draft=false,keepaspectratio=true,clip,width=\linewidth]{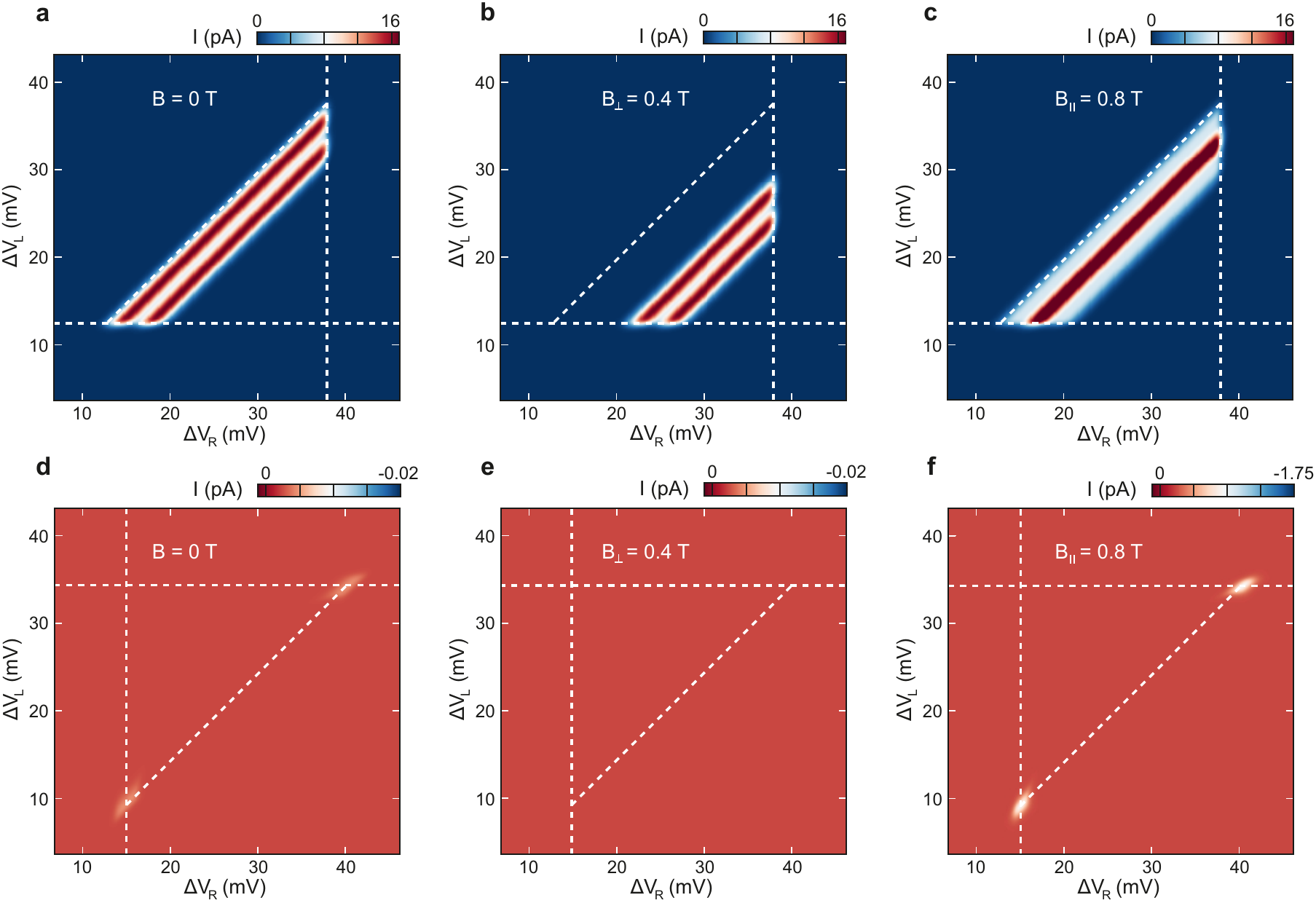}
\caption[S6]{  
Charge stability diagrams of the first triple point simulated by solving the rate equation. 
\textbf{a - c} depict the forward bias direction ($V_\mathrm{SD}=1$~mV) for different magnetic fields, showing the same features as the experimental data presented in Fig.~2. 
\textbf{d - f} show the blocked bias direction ($V_\mathrm{SD}=-1$~mV) for the same magnetic fields. For zero magnetic field, the blockade is lifted at the corners of the bias triangle, where back and forth tunneling to source (or drain) allows lifting the blockade. The effect is even larger at finite parallel magnetic fields, where the spins are tilted into the plane of the BLG.}
\label{S6}
\end{figure}

We follow this procedure for different magnetic fields and different gate voltage combinations $V_\text{L}$, $V_\text{R}$. The result is shown in Fig.~\ref{S6}, where we are able to reproduce the experimental data of Figs.~2b-d and Figs.~2f-h. Additionally, we simulate the current along the detuning axis of the $(0h,0e)\rightarrow(1h,1e)$ triple point as a function of parallel magnetic field, which is presented in Fig.~3e of the main manuscript.

\clearpage
\subsection{Electron-hole symmetry breaking due to Rashba spin-orbit coupling}
Since we are explicitly breaking the inversion symmetry of BLG with a perpendicular electric field, extrinsic (Rashba) spin-orbit coupling poses an additional mechanism to break the electron-hole symmetry in our DQD system. The corresponding full spin-orbit Hamiltonian acting on the low energy bands is then given by [6]
\begin{equation*}
\begin{split}
H_\mathrm{SO} =  \Psi^\dag \Big( & \frac{1}{4} [(\Delta^\text{t}_\mathrm{SO} + \Delta^\text{b}_\mathrm{SO}) \sigma_z - (\Delta^\text{t}_\mathrm{SO} -  \Delta^\text{b}_\mathrm{SO}) \sigma_0 ] \tau_z s_z  
 + \frac{1}{2} \lambda_\text{ex} (\sigma_y s_x + i \tau_z \sigma_x s_y ) \Big)  \Psi ,   
\end{split}
\end{equation*}
with the Pauli matrices $\tau, \sigma, s$ as defined in the main text, the extrinsic (Rashba) SO coupling  $\lambda_\text{ex}$, which scales linearly with the applied electric displacement field, and the proximity enhanced intrinsic (Kane-Mele) spin-orbit coupling energies $\Delta^\text{t}_\mathrm{SO}$ and $\Delta^\text{b}_\mathrm{SO}$ for the top and bottom layer of the BLG \footnote{$\Delta^\text{t}_\mathrm{SO}$, $\Delta^b_\mathrm{SO}$ and $\lambda_\text{ex}$ correspond to $\lambda_{I1}$ and $\lambda'_{I1}$ and $\lambda_3$ in Ref. \cite{Konschuh2012Mar}} \cite{Konschuh2012Mar}.
The influence of the proximity enhanced Kane-Mele spin-orbit coupling on electron-hole symmetry is discussed in the main text.

For understanding the influence of the extrinsic (Rashba) term, we note that for Fermi energies close to the band edge, the sublattice space is equivalent to the layer space and therefore to conduction and valence band. 
This is caused by the fact that excess charge is strongly layer polarized, only leading to a small admixture of the sublattices~\cite{McCann2013Apr, Banszerus2020May}.
The extrinsic SO term couples the two sublattices via $\sigma_{x,y}$ and therefore to the two layers, which experience a potential difference due to the electric displacement field.
As a consequence, the extrinsic spin-orbit term is suppressed to first order by~$\lambda_\mathrm{ex}^2/E_\mathrm{g}^2$.
Theoretical predictions of $\lambda_\text{ex}$ are at least three orders of magnitude smaller than the band gap ($E_\mathrm{g}$), rendering extrinsic spin-orbit coupling irrelevant for our system ~\cite{Konschuh2012Mar, Banszerus2021Sep}.

\clearpage
\subsection{Electron-hole symmetry breaking due to different valley g-factors in the electron and hole QDs}

\begin{figure}[!thb]
\centering
\includegraphics[draft=false,keepaspectratio=true,clip,width=0.8\linewidth]{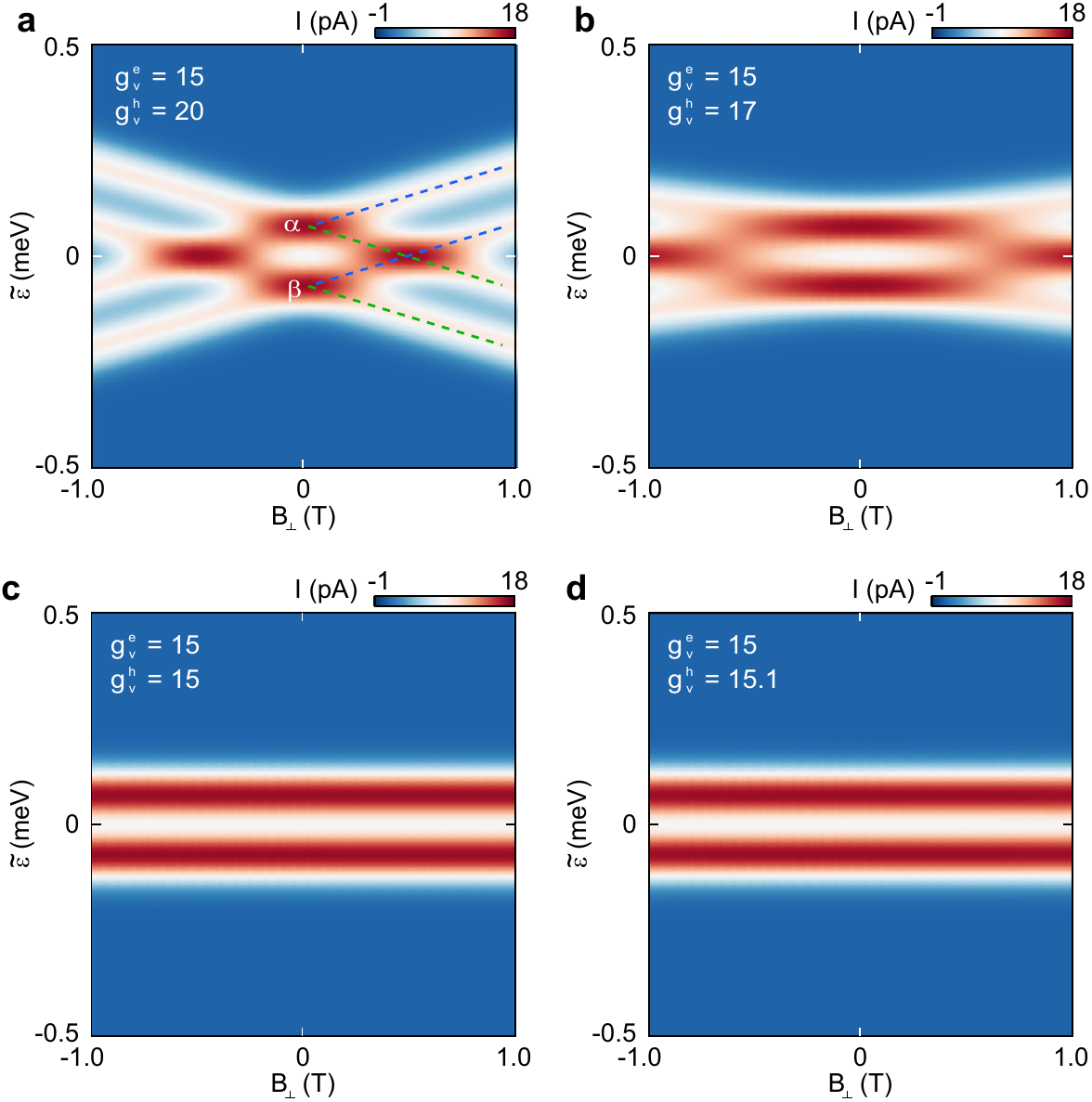}
\caption[Fig07]{Calculation of the current through the device as a function of the detuning energy $\widetilde \varepsilon$ (see arrow in Fig.~2c of the main text) and perpendicular magnetic field at a finite bias of $V_\mathrm{SD} = 1~$mV.
In \textbf{a}, the valley g-factors of the two QDs are chosen asymmetrically ($g^\text{e}_\text{v}=15$ for the electron QD and $g^\text{h}_\text{v}=20$ for the hole QD), resulting in a splitting of both, the $\alpha$ and $\beta$ transition, which scales with the difference in the valley g-factors. In \textbf{b}, the valley g-factors of the two QDs are chosen less asymmetrically ($g^\text{e}_\text{v}=15$ for the electron QD and $g^\text{h}_\text{v}=17$ for the hole QD), resulting in a smaller splitting of both, the $\alpha$ and $\beta$ transition, which scales with the difference in the valley g-factors. In \textbf{c} the valley g-factors are chosen symmetrically ($g_\text{v}=15$), and no dependence on $B_\perp$ is observed. In \textbf{d}, the experimentally observed g-factor difference of $g^\text{e}_\text{v}=15$ and $g^\text{h}_\text{v}=15.1$ is used for the simulation.
}
\label{S7}
\end{figure}

We investigate how asymmetric valley g-factors would affect the transition spectrum of the e-h DQD. In Fig.~\ref{S7}a-d we simulate the current through the device as a function of the detuning energy $\widetilde \varepsilon$ and perpendicular magnetic field, $B_\perp$, for different combinations of valley g-factors in the hole and electron QD, respectively. 
As clearly visible in Figs.~S7a,b, both the $\alpha$ and $\beta$ transition split due to the difference in valley g-factors (see colored lines in Fig.~\ref{S7}a) by $\Delta E = \frac{1}{2} \mu_\text{B} |g^\text{e}_\text{v}-g^\text{h}_\text{v}| B_\perp$. For equal valley g-factors, $\alpha$ and $\beta$ do not show any $B_\perp$-dependence, as shown in Fig.~S7c. A tiny asymmetry in valley g-factors is allowed without significantly changing the observed features for magnetic fields below 1T, as shown in Fig.~S7d, where a g-factor asymmetry of 0.1 is assumed.

\begin{figure}[!thb]
\centering
\includegraphics[draft=false,keepaspectratio=true,clip,width=0.85\linewidth]{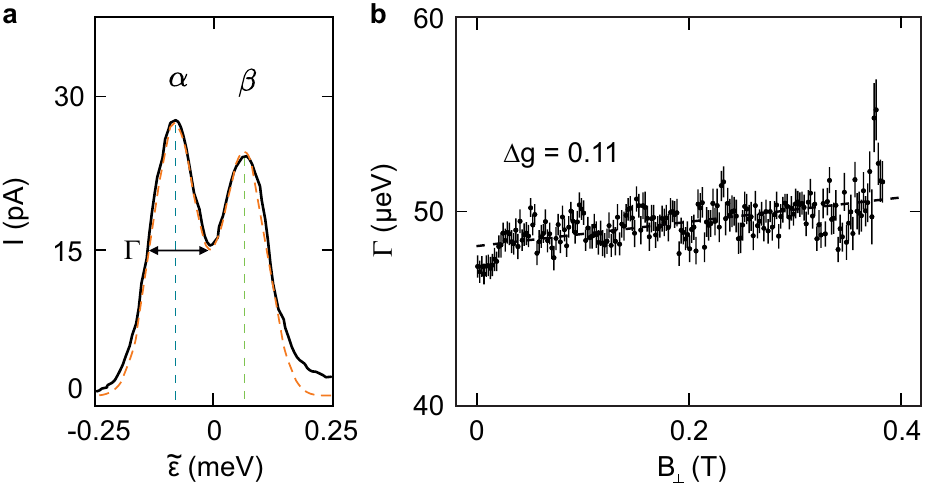}
\caption[Fig08]{\textbf{a} Exemplary line trace of the tunneling current as a function of the detuning. The sum of two Gauss curves with width $\Gamma$ is fitted to the data (see dashed line). \textbf{b}  $\Gamma$ extracted from the line fits as shown in a, as a function of $B_\perp$. Attributing the linear broadening of $\alpha$ and $\beta$ to an asymmetry of valley g-factors between electron and hole QD yields $\Delta g \approx 0.11$ .
}
\label{S8}
\end{figure}

To quantitatively estimate the valley g-factor asymmetry, we fit Gaussian peaks with width, $\Gamma$, to the detuning cuts presented in Fig.~3b in the main manuscript, allowing for a constant background and assuming equal width for both peaks, i.e. the $\alpha$ and $\beta$ peak. Such a fit is exemplarily shown in Fig.~S8a. The fitted width of the two peaks increases slightly for increasing $B_\perp$, as shown in Fig~\ref{S8}b. Attributing this effect entirely to a difference of the electron and hole g-factors, we obtain a maximum g-factor difference of $g_\text{v} \approx 0.1$ (c.f. with Fig.~S7d).

%